\documentclass[preprint,superscriptaddress]{revtex4-1} 

\usepackage{hyperref} 
\usepackage{amssymb}
\usepackage{amsmath}
\usepackage{graphicx}

\newcommand{\V}[1]{\mathbf{#1}}

\begin{document}

\title{Nonlocality in metallo-dielectric multilayers: numerical tools and physical analysis}

\author{Jessica Benedicto}
\affiliation{Clermont Universit\'e, Universit\'e Blaise
  Pascal, Institut Pascal, BP 10448, F-63000 Clermont-Ferrand, France}
\affiliation{CNRS, UMR 6602, IP, F-63171 Aubi\`ere, France} 
\author{R\'emi Poll\`es}
\affiliation{Clermont Universit\'e, Universit\'e Blaise
  Pascal, Institut Pascal, BP 10448, F-63000 Clermont-Ferrand, France}
\affiliation{CNRS, UMR 6602, IP, F-63171 Aubi\`ere, France} 
\author{Cristian Cirac\`i}
\affiliation{Instituto Italiano di Tecnologia (IIT), Center for Biomolecular Nanotechnologies, Via Barsanti, I-73010 Arnesano, Italy}
\affiliation{Center for Metamaterials and
  Integrated Plasmonics, Duke University, Durham, North Carolina
  27708, USA} 
\author{Emmanuel Centeno}
\affiliation{Clermont Universit\'e, Universit\'e Blaise
  Pascal, Institut Pascal, BP 10448, F-63000 Clermont-Ferrand, France}
\affiliation{CNRS, UMR 6602, IP, F-63171 Aubi\`ere, France} 
\author{David R. Smith}
\affiliation{Center for Metamaterials and
  Integrated Plasmonics, Duke University, Durham, North Carolina
  27708, USA} 
\author{Antoine Moreau}
\affiliation{Clermont Universit\'e, Universit\'e Blaise
  Pascal, Institut Pascal, BP 10448, F-63000 Clermont-Ferrand, France}
\affiliation{CNRS, UMR 6602, IP, F-63171 Aubi\`ere, France} 
\affiliation{Center for Metamaterials and
  Integrated Plasmonics, Duke University, Durham, North Carolina
  27708, USA} 

\begin{abstract}
We provide theoretical and numerical tools to quantitatively study 
the impact of nonlocality arising from 
free electrons in metals on the optical properties of metallo-dielectric
multilayers. Though effects due to nonlocality are in general quite small, they nevertheless can be important
for very thin (typically below 10 nm) metallic layers - as are used in structures characterized by
relatively flat dispersion curves. Such structures include those with negative refractive index; hyperbolic metamaterials; and materials with index near zero. We find in all cases that the inclusion of nonlocal effects through application of the hydrodynamic model to the electron response leads to
a higher transmission through the considered medium. Finally, we examine the excitation
of gap-plasmon resonances, where nonlocality plays a much greater role, and suggest possible routes for experimental investigation. 
\end{abstract}

\maketitle

Metallo-dielectric multilayers\cite{scalora98} are a class of
metamaterials that have attracted considerable attention because they are relatively easy to fabricate at infrared and visible wavelengths, and can support a
wide scope of exotic behavior, including negative
refraction and hyperbolic dispersion\cite{cai2005superlens,scalora07,hoffman2007negative,liu2007far,noginov2009bulk,cortes12,xu13}. The unique properties of these multilayer structures can potentially play a role in applications such as, thermal radiation
control\cite{biehs12,guo12,narimanov2012beyond} to subwavelength
imaging\cite{fang2005sub,belov05,cai2005superlens,jacob06,belov06,liu2007far,smolyaninov2007magnifying,zhang2008superlenses,mattiucci09,benedicto12}. Most of the analyses to date assume the classical, local model for the electron response of the metal layers. However, as the metals become very thin and are spaced close together, it can be expected that the classical model will no longer be applicable and must be substituted with a semiclassical or full quantum mechanical model. Recent experiments\cite{ciraci12} have indicated there may be a length scale where the classical model fails, but where the semiclassical, hydrodynamic model for the free electron response is valid. The hydrodynamic model, while approximate, nevertheless provides a more tractable model for analytic and semi-analytic calculations, especially for larger structures, and is thus an attractive approach for estimating the impact of nonlocality in various scenarios.

Over the years, there have been diverging opinions as to whether structures composed of very thin metallic and dielectric layers are sensitive to the intrinsic nonlocality of metals\cite{ruppin05,ruppin05b,yan13}, that results from the interaction between free electrons. Here we provide
a set of theoretical and numerical tools to take this nonlocality into account
when simulating the behavior of any kind of metallo-dielectric
multilayer. Our description relies on the hydrodynamic
model\cite{boardman82,forstmann86}, which has been shown to 
provide quantitative agreement for the plasmon resonance shifts observed on a system of gold nanospheres spaced sub-nanometer distances from a gold film 
\cite{ciraci12}. This model was subsequently modified to take into account interband transitions\cite{moreau13}.  
In this paper, we present a dispersion relation for an infinitely periodic metallo-dielectric
multilayer, as well as a scattering matrix formalism for finite layers, both of which incorporate the effects of nonlocality through the use of the hydrodynamic model. Our formalism makes it very easy to vary the boundary
conditions\cite{moreau13} and is thus more general than previous
works\cite{mochan87,yan12,yan13}.  Using these tools, we show that the
impact of nonlocality on a recently published optical negative index lens (n= -1) index lens design based on metallo-dielectric multilayers
\cite{xu13} can be observed but is very moderate. By contrast, we find nonlocal effects cannot be
neglected for structures in the canalization regime\cite{belov05};
moreover, for very thin metallic layers, the bulk plasmon acts as a
supplementary energy canal allowing light to tunnel through metallic layers.
Finally, we use our tools to study the prism-coupling of a 
gap-plasmon\cite{moreau13}, where the impact of nonlocality is found to be appreciable. Based on this observation, 
we are led to suggest an experimental configuration that could be used to quantitatively probe nonlocality in metals.

\section{Hydrodynamical model framework}

Within the framework of the hydrodynamical model, the conduction electrons of a metal are
treated as a gas trapped within a volume bounded by the surface of the metal structure. Interactions
between electrons are taken into account in an approximate manner through the introduction of a
pressure term that includes both the electrostatic and quantum pressure.
The hydrodynamical equations are linearized\cite{scalora10} to yield a relation
between the electric field and the polarization of the metal linked to
the free electrons displacements. The inclusion of the pressure terms in the hydrodynamic equations
introduces a spatial derivative into the linearized equation of motion, and the response of the electron gas is thus nonlocal\cite{crouseilles08}.

The origin of the hydrodynamic model can be traced to the physics of plasmas. It has been
used to describe metallic structures since the 1960s
and onwards\cite{kliewer68,melnyk70}, raising many questions along the way as to the proper treatment of a metal-dielectric interface. The model thus suffered from many
uncertainties, but was nonetheless widely
used\cite{fuchs81,gerhardts84,agarwal83,fuchs87} and
discussed\cite{boardman82,forstmann86,halevi95}, then somewhat abandoned -
likely due to the lack of any clear experimental evidence for nonlocal 
effects. It is clear now for instance, that nonlocality
has almost no impact on the surface plasmons of a thin metal film\cite{moreau13}, although the shift in the plasmon resonance wavelength of prism-coupled films
was often considered as a test for nonlocal
theories\cite{forstmann86}.

A variety of other approaches have been proposed to take nonlocality into account, including the classical Random Phase Approximation\cite{kliewer68,ford84,gerhardts84,chapuis08} or Feibelman's model\cite{feibelman82,liebsch1987dynamical}. It is interesting to note that an essential improvement to the latter approach concerns the manner in which the contribution from bound electrons is taken into account\cite{liebsch1995influence}. However, this model is know for not having issues to take bulk absorption into account\cite{chapuis08}.

More modern approaches to assessing electron response rely on density functional theory (DFT),  considered one of the most accurate tools to incorporate the effects of
quantum mechanical interactions on resonances and other properties of metallic nanoparticles
and nanoclusters\cite{savage12,teperik13}. These methods also have
limitations – for example, the propagation of waves is usually not 
taken directly into account\cite{esteban12}, which is a reasonable approximation for tightly confined structures such as spherical dimers,
but probably not descriptive for structures supporting gap-plasmons and similarly less confined excitations.

The hydrodynamical model has attracted increasing attention
due to advances in nanofabrication and measurement techniques that allow structures to be designed and studied in the regime where nonlocal response is expected to dominate \cite{ciraci2013hydrodynamic,luo13,david13}. In general, the hydrodynamic model is attractive because (i) it seems that its predictions are in good agreement with the first experiments for which
nonlocality clearly plays a role\cite{ciraci12}, and well before other
quantum effects kick in\cite{scholl12,scholl13}; (ii) it yields
analytical results\cite{fernandez12,moreau13} and provides deeper insight
into the physics of nonlocality; and (iii) it is easy to implement in
numerical simulations\cite{ciraci2013effects}. In addition, the
uncertainties about the boundary conditions are lifted when the
contribution of the bound and free electrons are clearly
distinguished\cite{moreau13} and its well-known tendency to
overestimate the impact of nonlocality is lessened. This is especially relevant
for very short wavelengths, when the permittivity is small and when the Drude term is of the order of the interband transition contribution. For the
reasons above, the model has thus been used to study 
the enhancement by plasmonics tips\cite{wiener12} and dimers
\cite{fernandez12a}, hyperbolic metamaterials\cite{yan12,yan13},
subwavelength imaging by a silver slab\cite{ruppin05,david13}
and gap-plasmon propagation\cite{moreau13,raza13,david13}.

It should be finally stressed even if the model was found to be
quite accurate\cite{ruppin05b} when compared to other
approaches, the model still has to be backed by more fundamental
studies\cite{teperik13}, comparison to experiments\cite{moreau12b}
or even more sophisticated models\cite{crouseilles08}.

We will in this section briefly remind the reader the fundamental
physics of nonlocality within the framework of the hydrodynamical model.

\subsection{Longitudinal waves}

\begin{figure}[h]
\begin{center}
\includegraphics[width=6cm]{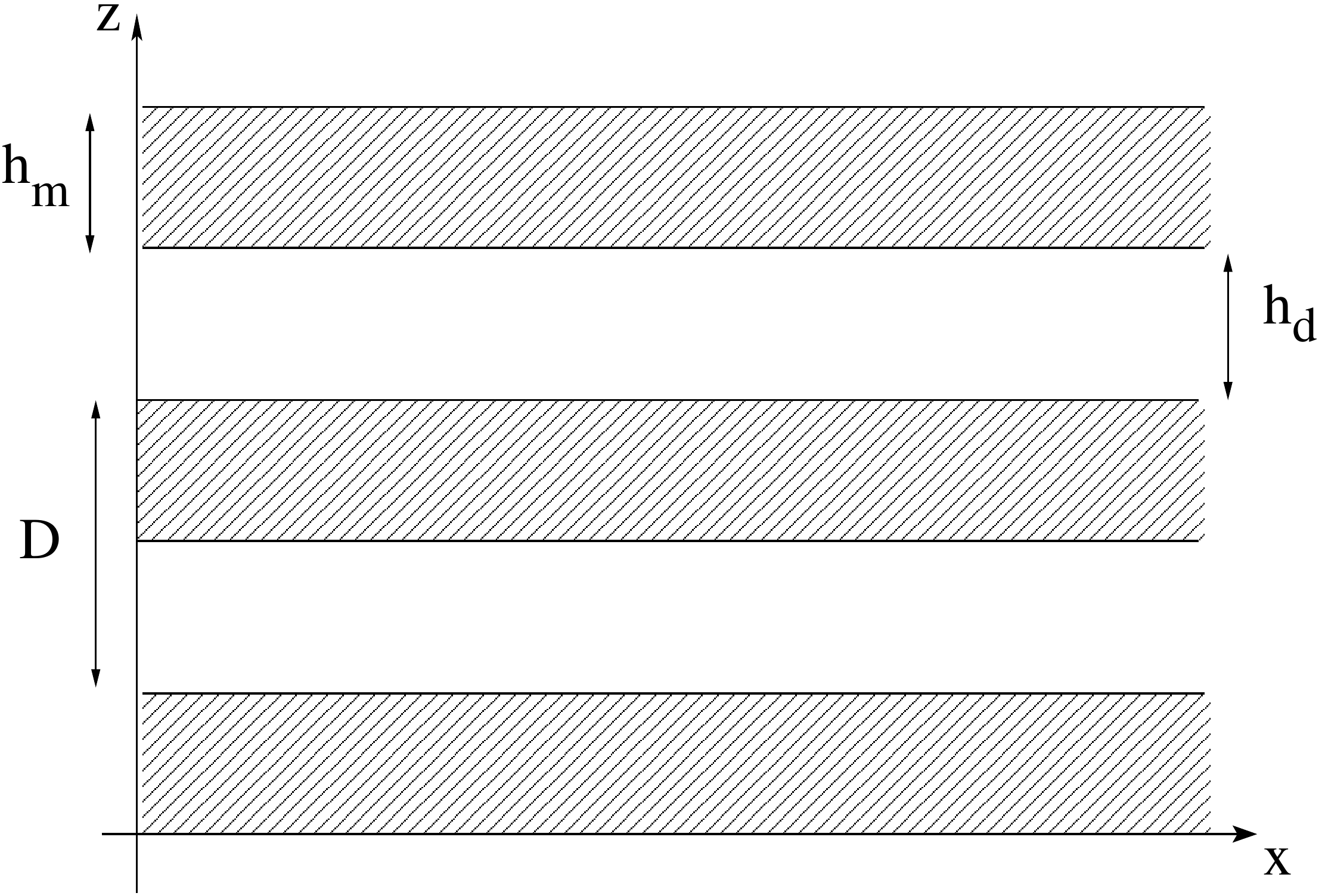}
\end{center}
\caption{Diagram of a periodic metallo-dielectric structure. The gray areas represent the metallic layers.
\label{fig:base}}
\end{figure}

We consider here a multilayer as represented in Fig. \ref{fig:base}. 
Nonlocal effects are expected to occur for $p$ polarization only (sometimes 
referred to as TM), so that we will from now on assume we are considering
this polarization only. We assume the structure is illuminated with a plane
wave, characterized by frequency $\omega$, a time dependency $e^{-i\omega t}$ 
and a wavevector whose component along the $x$ axis is denoted $k_x$.

In the framework of the hydrodynamic model\cite{raza11,ciraci13}, the electric and magnetic fields
satisfy Maxwell's equations
\begin{eqnarray}
 \nabla\times{\bf E}&=&i\omega\mu_0{\bf H}\\
 \nabla\times{\bf H}&=&-i\omega\epsilon_0(1+\chi_b){\bf E}+{\bf P}_f
\end{eqnarray}
where the effective polarization of the medium is linked to the electric field by the fundamental relation
\begin{equation}
{\bf P}_f=\frac{\epsilon_0.\omega^2_p}{\omega^2+i\gamma\omega}\left({\bf E}-(1+\chi_b)\frac{\beta^2}{\omega^2_p}\nabla\left(\nabla.{\bf E}\right)\right).
\end{equation}

These equations can be easily solved to yield an analytical form for the fields\cite{moreau13}. 
In the $j$th dielectric layer, having a relative permittivity $\epsilon_d$, the magnetic and electric fields can be written
\begin{eqnarray}
H_{y_d}&=&(A_j e^{i k_{d} z}+B_j e^{-i k_{d} z}) e^{i(k_x x-\omega t)}\\
E_{x}&=&\frac{ k_{d}}{\omega \varepsilon_{0} \varepsilon_{d}}(A_j e^{i k_{d} z}-B_j e^{-i k_{d} z}) e^{i(k_x x-\omega t)}\\
E_{z}&=&\frac{-k_x}{\omega \varepsilon_{0} \varepsilon_{d}}(A_j e^{i k_{d} z}+B_j e^{-i k_{d} z}) e^{i(k_x x-\omega t)},
\end{eqnarray}
with $k_d = \sqrt{\varepsilon_d k_0^2-k_x^2}$ and $k_0=\frac{\omega}{c}$.

Inside the $j$th metallic layer of permittivity $\varepsilon_m$, in the framework of the hydrodynamical model, two types of waves are supported: the transverse and longitudinal waves. The transverse wave results can be found from consideration of the two Maxwell curl equations. The associated transverse fields thus satisfy 
\begin{align}
\partial_z E_x -\partial_x E_z &= i\omega\,\mu_0 \,H_y\\ 
E_x &= \frac{1}{i\omega \varepsilon_0\,\varepsilon_m}\partial_z H_y\label{eq:ExHy}\\
E_z &=-\frac{1}{i\omega \varepsilon_0\,\varepsilon_m}\partial_x H_y.\label{eq:EzHy}
\end{align}
so that, inside the metal, they can be written
\begin{eqnarray}
H_{y_m}&=&(A_j e^{-\kappa_t z}+B_j e^{\kappa_t z}) e^{i(k_x x-\omega t)}\\
E^{t}_{x}&=&\frac{i \kappa_{t}}{\omega \varepsilon_{0} \varepsilon_{m}}(A_j e^{-\kappa_{t} z}-B_j e^{\kappa_{t} z}) e^{i(k_x x-\omega t)}\\
E^{t}_{z}&=&\frac{-k_x}{\omega \varepsilon_{0} \varepsilon_{m}}(A_j e^{-\kappa_{t} z}+B_j e^{\kappa_{t} z}) e^{i(k_x x-\omega t)},
\end{eqnarray}
where $\kappa_t = \sqrt{k_x^2-\varepsilon_m\,k_0^2}$.
The longitudinal wave corresponds to a bulk plasmon supported by the free electron gas, with no accompanying magnetic field. Because the electric field corresponding to the longitudinal mode is thus curl free, it satisfies
\begin{equation}
\partial_z E_x =\partial_x E_z \label{nulcurl}.
\end{equation}
and can finally be written (the first equation below being the definition of $C_j$ and $D_j$)
\begin{eqnarray}
E^{\ell}_{x}&=&\frac{1}{\omega \varepsilon_{0}}(C_j e^{-\kappa_{\ell} z}+D_j e^{\kappa_{\ell} z})\big] e^{i(k_x x-\omega t)}\\
E^{\ell}_{z}&=&\frac{-\kappa_{\ell}}{i k_x \omega \varepsilon_{0}}(C_j e^{-\kappa_{\ell} z}-D_j e^{\kappa_{\ell} z}) e^{i(k_x x-\omega t)}
\end{eqnarray}
with 
\begin{equation}
\kappa_\ell= \sqrt{k_x^2 + \frac{\omega_p^2}{\beta^2}\left(\frac{1}{\chi_f}+\frac{1}{1+\chi_b}\right)}
\end{equation}
where $\omega_p$ is the plasma frequency of the considered metal, and $\chi_f$ and $\chi_b$ are the susceptibilities associated with the free and bound electrons, respectively. These three parameters are determined through careful fits of the metal local permittivity\cite{rakic98}. The parameter $\beta$ is central, but not so easy to estimate. This constant can account for both coulomb interaction and quantum pressure through which free electrons interact strongly in the metal\cite{crouseilles08,scalora10}. The recent experimental results on film-coupled nanoparticles show consistency with the theoretically calculated value of $\beta=\sqrt{\frac{5}{3}\frac{E_F}{m}}$ (with $E_F$ being the Fermi level and $m$ the effective mass of free electrons in the metal\cite{crouseilles08}) \cite{ciraci12}. For gold and silver these values are very close, and we take here $\beta=1.35\times 10^6$ m/s for both metals.

\subsection{Additional boundary conditions}

At the boundary between a metal and a dielectric, an additional boundary condition is needed to determine to what extent the longitudinal wave is excited. Since 
we consider here the response of the free electrons as nonlocal and the response of the bound electrons as purely local, there is only one physically sound boundary
condition: We assume that the free electrons are confined within the metal volume, and thus the normal component of the polarization $\V{P}_f$ associated
with the free electron density vanishes at the interface\cite{moreau13}. We emphasize that this condition, although it may seem the most logical, has rarely been applied in prior
work\cite{mochan87,mochan88}. 

For the fields that are inside the metal, the boundary condition on the polarization can be written on the interface
\begin{equation}
  P_{f_z}=-\frac{1}{i \omega} \partial_x H_{y} - \epsilon_0 (1+\chi_b) E_{z} = 0
\end{equation}
where, of course, $\V{E}=\V{E}^t+\V{E}^\ell$. Using equation \ref{eq:EzHy}, the previous condition can also be written
\begin{equation}
  E^{\ell}_{z}=\frac{\kappa_{\ell}}{\omega \epsilon_0 k_x}  \Omega   H_{y}\label{eq:abc}
\end{equation}
with 
\begin{equation}
  \Omega = \frac{k_x^2}{\kappa_l}
  \left(
  \frac{1}{\epsilon}
  -
  \frac{1}{1+\chi_b}
  \right)\label{eq:Omega}.
\end{equation}

\section{Dispersion relation in an infinite metallo-dielectric multilayer}

Metallo-dielectric multilayers are of heightened interest in many metamaterial configurations for their unique dispersion characteristics. While fully isotropic negative index and other novel metamaterial medium are difficult to realize at visible and infrared wavelengths, anisotropic media comprising alternating metal/dielectric layers can be readily fabricated and can often approximate the desired wave propagation effects. In particular, one striking property that can be achieved in metallodielectric multilayers is hyperbolic dispersion\cite{smith03,smith04}. Hyperbolic metamaterials are compelling because the hyperbolic dispersion relation allows evanescent waves emitted by a source to be converted to propagating waves. This effect can be used to achieve sub-wavelength imaging\cite{belov05,belov06,bloemer07,mattiucci09,benedicto12,benedicto13}, projection of near-field information to far-field\cite{jacob06,liu2007far} and super-planckian thermal emission\cite{biehs12,narimanov2012beyond}. Analytically, the dispersion relation when metals are assumed to be local is extremely similar to the dispersion relation of Bragg mirrors. 

Several dispersion relations for {\em nonlocal} metallo-dielectric structures have already been published\cite{mochan87,yan13}, but do not consider the contribution of the interband transitions, which is expected to be quantitatively much more accurate when the correct boundary conditions are used.

Moreover, all the nonlocal parameters are united here in a single constant $\Omega$ making it very easy to change the characteristics of the model and thus allowing the retrieval of all the other dispersion relations. The demonstration leading to the dispersion relation is quite informative too. It shows not all the different transfer matrices algorithms\cite{krayzel10} can handle nonlocality easily.

\subsection{Dispersion relation}

We consider an infinite structure made of alternating metallic and dielectric layers with thicknesses of respectively $h_m$ and $h_d$. We will denote by a subscript $_m$ the fields inside the metal, and by $_d$ the fields inside the dielectric layer. 

The transfer matrix (Abel\`es matrices\cite{abeles50,markos08}) can be written

\begin{equation}
  \begin{bmatrix}
    H_{y_d}\\
    E_{x}
  \end{bmatrix}_{z+h_d}=
  \begin{bmatrix}
    \cos(k_d h_d)&\frac{i \omega \varepsilon_0 \varepsilon_{d}}{k_d} \sin(k_d h_d)\\
    \frac{i k_d}{\omega \varepsilon_0 \varepsilon_{d}} \sin(k_d h_d)&\cos(k_d h_d) 
  \end{bmatrix}
  \begin{bmatrix}
    H_{y_d}\\
    E_{x}
  \end{bmatrix}_{z}.
\end{equation}

which relates the transverse fields on either side of the dielectric layer.

In the metal, since both transverse as well as longitudinal waves propagate, two transfer matrices must be written, or 

\begin{equation}
  \begin{bmatrix}
    H_{y_m}\\
    E^{t}_{x}
  \end{bmatrix}_{z+h_m}=
  \begin{bmatrix}
    \cosh(\kappa_t h_m)&\frac{i \omega \varepsilon_0 \varepsilon_{m}}{\kappa_t} \sinh(\kappa_t h_m)\\
    \frac{\kappa_t}{i \omega \varepsilon_0 \varepsilon_{m}} \sinh(\kappa_t h_m)&\cosh(\kappa_t h_m)
  \end{bmatrix}
  \begin{bmatrix}
    H_{y_m}\\
    E^{t}_{x}
  \end{bmatrix}_{z}
\end{equation}

and

\begin{equation}
  \begin{bmatrix}
    E^{\ell}_{x}\\
    E^{\ell}_{z}
  \end{bmatrix}_{z+h_m}=
  \begin{bmatrix}
    \cosh(\kappa_\ell h_m)&\frac{i k_x}{\kappa_\ell} \sinh(\kappa_\ell h_m)\\
    \frac{\kappa_\ell}{i k_x} \sinh(\kappa_\ell h_m)&\cosh(\kappa_\ell h_m) 
  \end{bmatrix}
  \begin{bmatrix}
    E^{\ell}_{x}\\
    E^{\ell}_{z}
  \end{bmatrix}_{z}
\end{equation} 

For a given period, writing down all the boundary conditions leads to the following equations:

The continuity of $E_x$ yields
\begin{eqnarray}
  E_{x} (0) &=& E^{\ell}_{x} (0)  +  E^{t}_{x} (0)\\	
  E_{x} (h_m) &=& E^{\ell}_{x} (h_m)  +  E^{t}_{x} (h_m),
\end{eqnarray}
while the continuity of $H_y$ gives
\begin{eqnarray}
  H_{y_d}(0)&=&H_{y_m}(0)\\
  H_{y_d}(h_m)&=&H_{y_m}(h_m).
\end{eqnarray}
Finally, if we apply the additional boundary conditions discussed above, we get the supplementary equations
\begin{eqnarray}
E^{\ell}_{z} (0) = \frac{\kappa_{\ell}}{\omega \epsilon_0 k_x}  \Omega   H_{y_m} (0) \\
E^{\ell}_{z} (h_m) = \frac{\kappa_{\ell}}{\omega \epsilon_0 k_x}  \Omega  H_{y_m} (h_m).
\end{eqnarray}

All the above equations (the transfer matrixes and the ones produced by the boundary conditions) can be used to yield

\begin{equation}
\begin{bmatrix}
    H_{y_d}\\
    E_{x} 
\end{bmatrix}
_{h_m+h_d}=
\begin{bmatrix}
    C_{11} & C_{12}\\
    C_{21} & C_{22}
\end{bmatrix}
\begin{bmatrix}
    H_{y_d} \\
    E_{x}
\end{bmatrix}_{0}
\end{equation}

where the coefficients of the matrix, denoted $C$, are 
\begin{widetext}
\begin{eqnarray}
C_{11} &=&\left[\cos(k_d h_d) \cosh(\kappa_t h_m) +\left(\frac{\varepsilon_{d}}{k_d} \frac{\kappa_t}{\varepsilon_{m}} +\frac{\varepsilon_{d}}{k_d} \frac{\varepsilon_{m}}{\kappa_t} \Omega^2\right)\sin(k_d h_d) \sinh(\kappa_t h_m) \right.\nonumber\\
&&-\frac{\Omega \varepsilon_{m}}{\kappa_t} \cos(k_d h_d) \cosh(\kappa_\ell h_m)\frac{\sinh(\kappa_t h_m)}{\sinh(\kappa_\ell h_m)}\nonumber\\
 & &\left.+\frac{\Omega \varepsilon_{d}}{k_d} \frac{\sin(k_d h_d)}{\sinh(\kappa_\ell h_m)} \left(2 -2 \cosh(\kappa_t h_m) \cosh(\kappa_\ell h_m)\right)\right] \left(1 -\frac{\Omega \varepsilon_{m}}{\kappa_t} \frac{\sinh(\kappa_t h_m)}{\sinh(\kappa_\ell h_m)}\right)^{-1}
\end{eqnarray}

\begin{eqnarray}
  C_{12} &=&\left(\cos(k_d h_d) -\frac{\frac{\Omega \varepsilon_{d}}{k_d} \left[\cosh(\kappa_\ell h_m)-\cosh(\kappa_t h_m)\right] \sin(k_d h_d)}{\sinh(\kappa_\ell h_m)}\right)\left(\frac{\frac{i \omega \varepsilon_0 \varepsilon_{m}}{\kappa_t} \sinh(\kappa_t h_m)}{1 -\frac{\varepsilon_{m}}{\kappa_t} \Omega \frac{\sinh(\kappa_t h_m)}{\sinh(\kappa_\ell h_m)}}\right)\nonumber\\
  &&+\frac{i \omega \varepsilon_0 \varepsilon_{d}}{k_d} \sin(k_d h_d)	\cosh(\kappa_t h_m)
\end{eqnarray}

\begin{eqnarray}
  C_{21} &=&\left(\frac{i k_d}{\omega \varepsilon_0 \varepsilon_{d}} \sin(k_d h_d) +\frac{\frac{i \Omega}{\omega \varepsilon_{0}} [\cosh(\kappa_\ell h_m)-\cosh(\kappa_t h_m)] \cos(k_d h_d)}{\sinh(\kappa_\ell h_m)}\right)\nonumber\\
  &&\left(\frac{\cosh(\kappa_t h_m) -\frac{\Omega \varepsilon_{m}}{\kappa_t} \sinh(\kappa_t h_m) \tanh(\kappa_\ell h_m)}{1 -\frac{\Omega \varepsilon_{m}}{\kappa_t} \frac{\sinh(\kappa_t h_m)}{\sinh(\kappa_\ell h_m)}}\right)\nonumber\\
  &&+ \frac{\kappa_t}{i \omega \varepsilon_{0} \varepsilon_{m}} \cos(k_d h_d) \sinh(\kappa_t h_m) +\frac{i \Omega}{\omega \varepsilon_{0}} \cos(k_d h_d) \sinh(\kappa_\ell h_m) \nonumber\\
  &&-\frac{\frac{i \Omega}{\omega \varepsilon_{0}} [\cosh(\kappa_\ell h_m)-\cosh(\kappa_t h_m)] \cosh(\kappa_\ell h_m) \cos(k_d h_d)}{\sinh(\kappa_\ell h_m)}
\end{eqnarray}
and finally
\begin{eqnarray}
C_{22} &=&\left(\frac{i k_d}{\omega \varepsilon_0 \varepsilon_{d}} \sin(k_d h_d) +\frac{\frac{i \Omega}{\omega \varepsilon_{0}} [\cosh(\kappa_\ell h_m)-\cosh(\kappa_t h_m)] \cos(k_d h_d)}{\sinh(\kappa_\ell h_m)}\right)\left(\frac{\frac{i \omega \varepsilon_0 \varepsilon_{m}}{\kappa_t} \sinh(\kappa_t h_m)}{1 -\frac{\varepsilon_{m}}{\kappa_t} \Omega \frac{\sinh(\kappa_t h_m)}{\sinh(\kappa_\ell h_m)}}\right)\nonumber\\
&&+\cos(k_d h_d) \cosh(\kappa_t h_m)
\end{eqnarray}
\end{widetext}

We seek solutions that correspond to propagating modes in the structure, and that are expected to be pseudo-periodic, and thus satisfy the Bloch-Floquet condition
\begin{equation}
\begin{bmatrix}
  H_{y_d}\\
  E_{x} 
\end{bmatrix}_{h_m+h_d}=
e^{\pm i K D}
\begin{bmatrix}
  H_{y_d} \\
  E_{x}
\end{bmatrix}_{0}
\end{equation}
The two solutions can be either propagating upward or downward, and $e^{\pm i K D}$
appear as the two eigenvalues of the matrix $C$, and thus as the two solutions of the characteristic equation
\begin{equation}
  \lambda^2-Tr(C) \lambda+Det(C)=0
\end{equation}
which means we have 
\begin{equation}
e^{\pm i K D}=\frac{1}{2}\left[Tr(C)\pm\sqrt{Tr(C)^2-4 Det(C)}\right].
\end{equation}
Another way to write what is already a dispersion relation, is to sum up the two eigenvalues to get 
\begin{equation}
  \cos(K D)=\frac{1}{2} (C_{11}+C_{22}),
\end{equation}
which can be written, using the previous expressions of $C_{11}$ and $C_{22}$, 
\begin{widetext}
\begin{eqnarray}
&\left(1 -\frac{\Omega \varepsilon_{m}}{\kappa_t} \frac{\sinh(\kappa_t h_m)}{\sinh(\kappa_\ell h_m)}\right)cos(K D)=\nonumber\\
&\cos(k_d h_d) \cosh(\kappa_t h_m) +\frac{1}{2}\left(\frac{\varepsilon_{d}}{k_d} \frac{\kappa_t}{\varepsilon_{m}} - \frac{k_d}{\varepsilon_{d}} \frac{\varepsilon_{m}}{\kappa_t} +\frac{\varepsilon_{d}}{k_d} \frac{\varepsilon_{m}}{\kappa_t} \Omega^2\right)\sin(k_d h_d) \sinh(\kappa_t h_m)\nonumber\\
&+\frac{\Omega}{\sinh(\kappa_\ell h_m)}\left[\frac{\varepsilon_{d}}{k_d} \sin(k_d h_d) \left(1 -\cosh(\kappa_t h_m) \cosh(\kappa_\ell h_m)\right)-\frac{\varepsilon_{m}}{\kappa_t} \cos(k_d h_d) \cosh(\kappa_\ell h_m) \sinh(\kappa_t h_m)\right]\label{e:rdisp}
\end{eqnarray}
\end{widetext}

This expression can be checked for consistency in various limits. For instance, making $\Omega$ vanish reduces the expression to the local dispersion relation for a hyperbolic structure. When the metal thickness tends to infinity, the dispersion relation reduces to the gap-plasmon dispersion relation\cite{moreau13,raza13}. Furthermore, this expression recovers all previously published results with different boundary conditions\cite{mochan87,yan13} by just changing the expression of $\Omega$ and $\kappa_l$, as done in \cite{moreau13}.

\section{Scattering matrix algorithm}

Even if the previous dispersion relation is extremely informative and allows to retrieve very important parameters like the effective index\cite{benedicto12} for metallo-dielectric structures, structures with a more complex pattern have recently attracted a lot of attention\cite{verhagen10,xu13}. We underline that even for structures with simple patterns, the first and the last layer are usually made of the same material thus limiting the insight the dispersion relation can provide : it does not give access to the reflection coefficient\cite{yan13}. For these reasons, a systematic way for calculating the reflection coefficient of a metallo-dielectric structure and the field inside any layer is required. This is all the more necessary that cavity resonances in metallo-dielectric layers\cite{benedicto11b} may complicate the global picture given by the dispersion curves.

A transfer matrix method has been proposed in the eighties\cite{mochan88} to study periodical multilayers above the plasma frequency. But transfer matrices are not numerically stable, especially below the plasma frequency when the waves are evanescent\cite{krayzel10}, and are difficult to use systematically. As we will see in this section, the scattering matrices are adapted to taking nonlocality into account and they are numerically perfectly stable so that they constitute a natural choice. Our method has been validated by comparison with full COMSOL simulations, especially for the gap-plasmon resonance excitation.

\subsection{Layer matrices}

We consider a layer $j$, comprised between two interfaces located at $z=z_j$ for the upper interface and $z=z_{j+1}$ for the lower one. The thickness of the layer is $h_j=z_j-z_{j+1}$. It is convenient to introduce here the coefficients $A_j^\pm$ and $B_j^\pm$, that are defined in a dielectric layer by
\begin{eqnarray}
H_y &=& \left(A^+_j e^{i k_z^j (z-z_j)}+B^+_j e^{-i k_z^j (z-z_j)}\right) e^{i(k_x x-\omega t)}\\
&=& \left(A^-_j e^{i k_z^j (z-z_{j+1})}+B^-_j e^{-i k_z^j (z-z_{j+1})}\right) e^{i(k_x x-\omega t)}
\end{eqnarray}
with $k_z^j = \sqrt{\varepsilon_j k_0^2-k_x^2}$, $\varepsilon_j$ being the relative permittivity of the dielectric medium. Using Maxwell's equations for $p$ polarisation \eqref{eq:EzHy}, the electric field ($E_x$,$E_z$) can be easily calculated.

This leads to introducing a scattering matrix for a dielectric layer that writes
\begin{equation}
\begin{bmatrix}
A_j^+\\ B_j^-
\end{bmatrix}
=\begin{bmatrix}
0 & e^{i k_z^j \,h_j}\\
e^{i k_z^j \,h_j} & 0
\end{bmatrix}
\begin{bmatrix}
B_j^+\\ A_j^-
\end{bmatrix}.
\end{equation}

Inside a metallic layer, coefficients can similarly be defined for the transversal wave as
\begin{eqnarray}
H_y &=&  \left(A^+_j e^{-\kappa_t (z-z_j)}+B^+_j e^{\kappa_t (z-z_j)}\right) e^{i(k_x x-\omega t)}\\
&=&\left(A^-_j e^{-\kappa_t (z-z_{j+1})}+B^+_j e^{\kappa_t (z-z_{j+1})}\right) e^{i(k_x x-\omega t)},
\end{eqnarray}
and the corresponding electric field can be determined using \eqref{eq:EzHy}. Taking the longitudinal wave into account leads to introducing coefficients $C_j^\pm$ and $D_j^\pm$:
\begin{eqnarray}
E^{\ell}_{x}&=&\frac{1}{\omega \varepsilon_{0}}(C^{+}_{j} e^{-\kappa_{\ell} (z-z_j)}+D^{+}_{j} e^{\kappa_{\ell} (z-z_{j})}) e^{i(k_x x-\omega t)}\\
&=&\frac{1}{\omega \varepsilon_{0}}(C^{-}_{j} e^{-\kappa_{\ell} (z-z_{j+1})}+D^{-}_{j} e^{\kappa_{\ell} (z-z_{j+1})}) e^{i(k_x x-\omega t)}.
\end{eqnarray}

This leads to a scattering matrix for a metallic layer that writes
\begin{equation}
  \begin{bmatrix}
    A^{+}_j\\
    C^{+}_j\\
    B^{-}_j\\
    D^{-}_j 
  \end{bmatrix}	=
  \begin{bmatrix}
    0 & 0 & e^{- \kappa_t h_j} & 0\\
    0 & 0 & 0 & e^{- \kappa_{\ell} h_j}\\
    e^{- \kappa_{t} h_j} & 0 & 0 & 0\\
    0 & e^{- \kappa_{\ell} h_j} & 0 &0 
  \end{bmatrix}
  \begin{bmatrix}
    B^{+}_j\\
    D^{+}_j\\
    A^{-}_j\\
    C^{-}_j
  \end{bmatrix}.
\end{equation}

\subsection{Dielectric to metal scattering matrix}

We assume here that medium $j$ is dielectric, while medium $j+1$ is
metallic. At such an interface, the magnetic field $H_y$ is continuous,
as is $E_x$, that can be calculated using \eqref{eq:ExHy}. A supplementary condition
inside the metal is given by \eqref{eq:abc}. A straightforward calculation shows that
this leads to the following conditions on the coefficients
\begin{eqnarray}
  A^-_j+B^-_j &=& A^{+}_{j+1}+B^{+}_{j+1}\label{eq:contH}\\
  \frac{k_z^j}{\epsilon_j} (A^-_j - B^-_j) &=& \frac{i \kappa_{t}^{j+1}}{\epsilon_{j+1}} (A^{+}_{j+1} - B^{+}_{j+1})\\
&&\nonumber  + C^{+}_{j+1} + D^{+}_{j+1}\label{eq:contE}\\	
  D^{+}_{j+1} - C^{+}_{j+1} &=& i \Omega (A^{+}_{j+1} + B^{+}_{j+1})\label{eq:contP}.
\end{eqnarray}

Rearranging these equations, a scattering matrix for the dielectric-metal interface can be written 
\begin{widetext}
\begin{equation}
\begin{bmatrix}
A_j^-\\ B^+_{j+1} \\ D_{j+1}^+
\end{bmatrix}=
\frac{1}{\frac{k_z^j}{\epsilon_j}+\frac{i \kappa_{t}^{j+1}}{\epsilon_{j+1}} - i \Omega}
\begin{bmatrix}
  \frac{k_z^j}{\epsilon_j}- \frac{i \kappa_{t}^{j+1}}{\epsilon_{j+1}} + i \Omega & 2 \frac{i \kappa_{t}^{j+1}}{\epsilon_{j+1}} & 2\\
  2\frac{k_z^j}{\epsilon_j} & \frac{i \kappa_{t}^{j+1}}{\epsilon_{j+1}} -\frac{k_z^j}{\epsilon_j}+i\Omega & 2\\
  2 i \Omega \frac{k_z^j}{\epsilon_j} & 2 i \Omega \frac{i \kappa_{t}^{j+1}}{\epsilon_{j+1}} & \frac{k_z^j}{\epsilon_j} + \frac{i \kappa_{t}^{j+1}}{\epsilon_{j+1}}+i \Omega
\end{bmatrix}
\begin{bmatrix}
B_j^-\\ A^+_{j+1} \\ C_{j+1}^+
\end{bmatrix}
\end{equation}
\end{widetext}

\subsection{Metal to dielectric scattering matrix}

Similarly, at the interface between a metal (upper layer $j$) and a dielectric (lower layer $j+1$), the boundary conditions lead to the following equations
\begin{eqnarray}
  A^-_j+B^-_j &=& A^{+}_{j+1}+B^{+}_{j+1}\label{eq:contH2}\\
  \frac{i\:\kappa_{t}^j}{\epsilon_j} (A^-_i - B^-_i) + C^{-}_{j} + D^{-}_{j} &=& \frac{k_z^{j+1}}{\epsilon_{j+1}} (A^{+}_{j+1} - B^{+}_{j+1})\label{eq:contE2}\\	
  D^{-}_{j} - C^{-}_{j}&=&i\Omega\,(A^{-}_{j} + B^{-}_{j})\label{eq:contP2},
\end{eqnarray}
that become, once they have been re-arranged,
\begin{widetext}
\begin{equation}
  \begin{bmatrix}
    A_j^- \\ C_j^- \\ B_{j+1}^+
  \end{bmatrix}
=
  \frac{1}{\frac{i\:\kappa_{t}^j}{\epsilon_j}\:+\:\frac{k_z^{j+1}}{\epsilon_{j+1}}\:-\:i\:\Omega}
  \begin{bmatrix}
    \frac{i\:\kappa_{t}^j}{\epsilon_j}\:-\:\frac{k_z^{j+1}}{\epsilon_{j+1}}\:+\:i\:\Omega & -\:2 & 2\:\frac{k_z^{j+1}}{\epsilon_{j+1}}\\
    -\:2\:i\:\Omega\:\frac{i\:\kappa_{t}^j}{\epsilon_j}  & \frac{i\:\kappa_{t}^j}{\epsilon_j}\:+\:\frac{k_z^{j+1}}{\epsilon_{j+1}}\:+\:i\:\Omega & -\:2\:i\:\Omega\:\frac{k_z^{j+1}}{\epsilon_{j+1}}\\
    2\:b^-_{i} & -\:2 & \frac{k_z^{j+1}}{\epsilon_{j+1}}\:-\:\frac{i\:\kappa_{t}^j}{\epsilon_j}\:+\:i\:\Omega
  \end{bmatrix}
  \begin{bmatrix}
        B_j^- \\ D_j^- \\ A_{j+1}^+
  \end{bmatrix}
\end{equation}
\end{widetext}

\subsection{Cascading method}

Now that scattering matrixes have been defined for all kinds of interfaces, they have to be combined through a cascading method. Here the scattering matrixes are all square matrixes, but they may have more than 2 lines, so that the usual cascading algorithm cannot be used. Instead, one has to rely on the cascading algorithm that is usually employed for Fourier Modal Methods\cite{lalanne96,granet96}. When trying to combine a scattering matrix $\V{S}$ so that 

\begin{equation}
\begin{bmatrix}
\V{A}\\ \V{B}
\end{bmatrix}
=
\begin{bmatrix}
 \V{S}_{11} &  \V{S}_{12}\\  \V{S}_{21} &  \V{S}_{22}
\end{bmatrix}
\begin{bmatrix}
\V{C}\\ \V{D}
\end{bmatrix}
\end{equation}
with a scattering matrix $\V{U}$ such that
\begin{equation}
\begin{bmatrix}
\V{D}\\ \V{E}
\end{bmatrix}
=
\begin{bmatrix}
\V{U}_{11} & \V{U}_{12}\\ \V{U}_{21} & \V{U}_{22}
\end{bmatrix}
\begin{bmatrix}
\V{B}\\ \V{F}
\end{bmatrix}
\end{equation}
then the resulting scattering matrix is
\begin{widetext}
\begin{equation}
\begin{bmatrix}
\V{A}\\ \V{E}
\end{bmatrix}
=
\begin{bmatrix}
\V{S}_{11} + \V{S}_{12}(\V{1}-\V{S}_{11}\V{U}_{22})^{-1} \V{U}_{11} \V{S}_{21} &
\V{S}_{12}(\V{1}-\V{S}_{11}\V{U}_{22})^{-1} \V{U}_{12}\\
\V{U}_{21}(1-\V{S}_{22}\V{U}_{11})^{-1} \V{S}_{21} & 
\V{U}_{22}+\V{U}_{21}(1-\V{S}_{22}\V{U}_{11})^{-1} \V{S}_{22}\V{U}_{12}
\end{bmatrix}
\begin{bmatrix}
\V{C}\\ \V{F}
\end{bmatrix}.
\end{equation}
\end{widetext}

This method can be applied here, even if the $\V{U}_{ij}$ are in general not square. Here
$\V{A}, \V{B}, \V{C}, \V{D}, \V{E}$ and $\V{F}$ may represent vectors as $\begin{bmatrix} A_j^\pm\\C_j^\pm\end{bmatrix}$, $\begin{bmatrix} B_j^\pm\\D_j^\pm\end{bmatrix}$, or simply $\begin{bmatrix} A_j^\pm \end{bmatrix}$ or $\begin{bmatrix} B_j^\pm \end{bmatrix}$ depending on the scattering matrix. Each time a cascade is needed, there is however no ambiguity on how to choose the vectors, given the size of the matrixes that have to be cascaded.

\begin{widetext}
In order to compute the field inside the layers, beyond the reflection and transmission coefficients of the whole structure, it is necessary to compute the vectors that are eliminated during the cascading process. They can be obtained through the following relations
\begin{equation}
\begin{bmatrix}
\V{B}\\ \V{D}
\end{bmatrix}
=
\begin{bmatrix}
(1-\V{S}_{22}\V{U}_{11})^{-1} \V{S}_{21}&
(1-\V{S}_{22}\V{U}_{11})^{-1} \V{S}_{22}\V{U}_{12}\\
(\V{1}-\V{S}_{11}\V{U}_{22})^{-1} \V{U}_{11} \V{S}_{21} &
(\V{1}-\V{S}_{11}\V{U}_{22})^{-1} \V{U}_{12}
\end{bmatrix}
\begin{bmatrix}
\V{C}\\ \V{F}
\end{bmatrix}.
\end{equation}
\end{widetext}

Finally, once all the $A_j^\pm$, $B_j^\pm$ have been obtained using the previous method, the most stable way to compute the magnetic field (but this is true for any other field) is to use the following hybrid expression inside a layer
\begin{equation}
H_y = \left(A_j^- e^{ik_z^j (z-z_{j+1})}+B_j^+ e^{-ik_z^j (z-z_j)}\right)\,e^{i(k_x x - \omega t)}
\end{equation}
with $k_z^j= i\kappa_t^j$ in the case of a metallic layer, to ensure the exponentials have a modulus as small as possible.


\section{Impact metamaterials based on metallo-dielectric multilayers}

\subsection{Hyperbolic dispersion relation}

We now study the impact of nonlocality on the dispersion relation \eqref{e:rdisp} in the case of a hyperbolic dispersion relation from a numerical point of view. One of the most promising experimental realizations of a medium with an engineered dispersion relation is the n=-1 effective index medium fabricated with alternating layers of silver and titanium oxide\cite{xu13}, at an operating wavelength of $363.8$ nm. We choose here to consider the same kind of structures, especially the same media and properties, in order to be as realistic as possible.

\begin{figure}[h]
\begin{center}
\includegraphics[width=8cm]{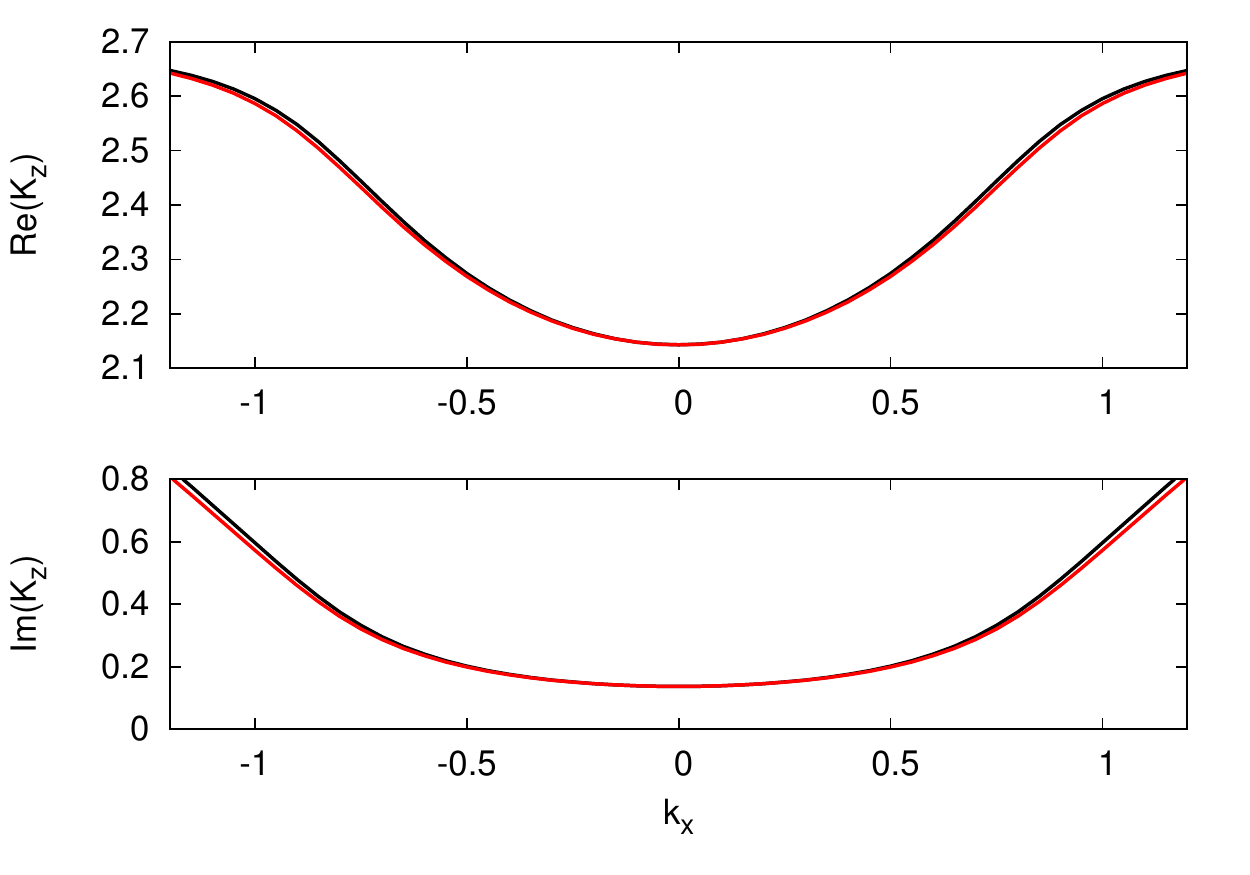}
\end{center}
\caption{Dispersion curves for a metallo-dielectric structure ($h_d=33.7$ nm, $h_m=28.1$ nm, $\lambda=363.8$ nm) presenting an effective index close to -1. Top: Real part. Bottom: Imaginary part. The local (black line) and nonlocal (red line) cases are shown. Very little difference can be seen, especially close to normal incidence.
\label{fig:disp1}}
\end{figure}

Figure \ref{fig:disp1} shows the dispersion relation for an hyperbolic isofrequency curve presenting an effective index\cite{benedicto12,benedicto13} close to $-1$. Obviously, the impact of nonlocality is negligible on this kind of structure. The change in the curvature of the dispersion relation induces a small change in the effective index of the structure, which may in turn cause a small change in the position of the image given by the lens equation\cite{benedicto12}. But the change seems very small compared to what is suggested in previous works\cite{yan13}.

Nonlocal effects are expected for much thinner metallic layers that those that are considered here with a thickness of $28.1$ nm for metallic layers. In order for these effects to have an impact, the thickness of the layer needs not to be much larger than the penetration length of the longitudinal waves in the metal. 

It turns out that, keeping the same period, the metallic layers are much thinner for a larger effective index (corresponding to a smaller curvature\cite{benedicto12,benedicto13}). In addition, when the dispersion curve is flattened it is much more sensitive to any perturbation. These are the reasons why nonlocality has a much larger impact on structures in the canalization regime\cite{belov05,belov06} where metallic layers are much thinner than the dielectric ones. In order to illustrate this idea, we choose to study a periodic medium alterning a 5 nm metallic layer with a 22.6 nm dielectric layer.

\begin{figure}[h]
\begin{center}
\includegraphics[width=8cm]{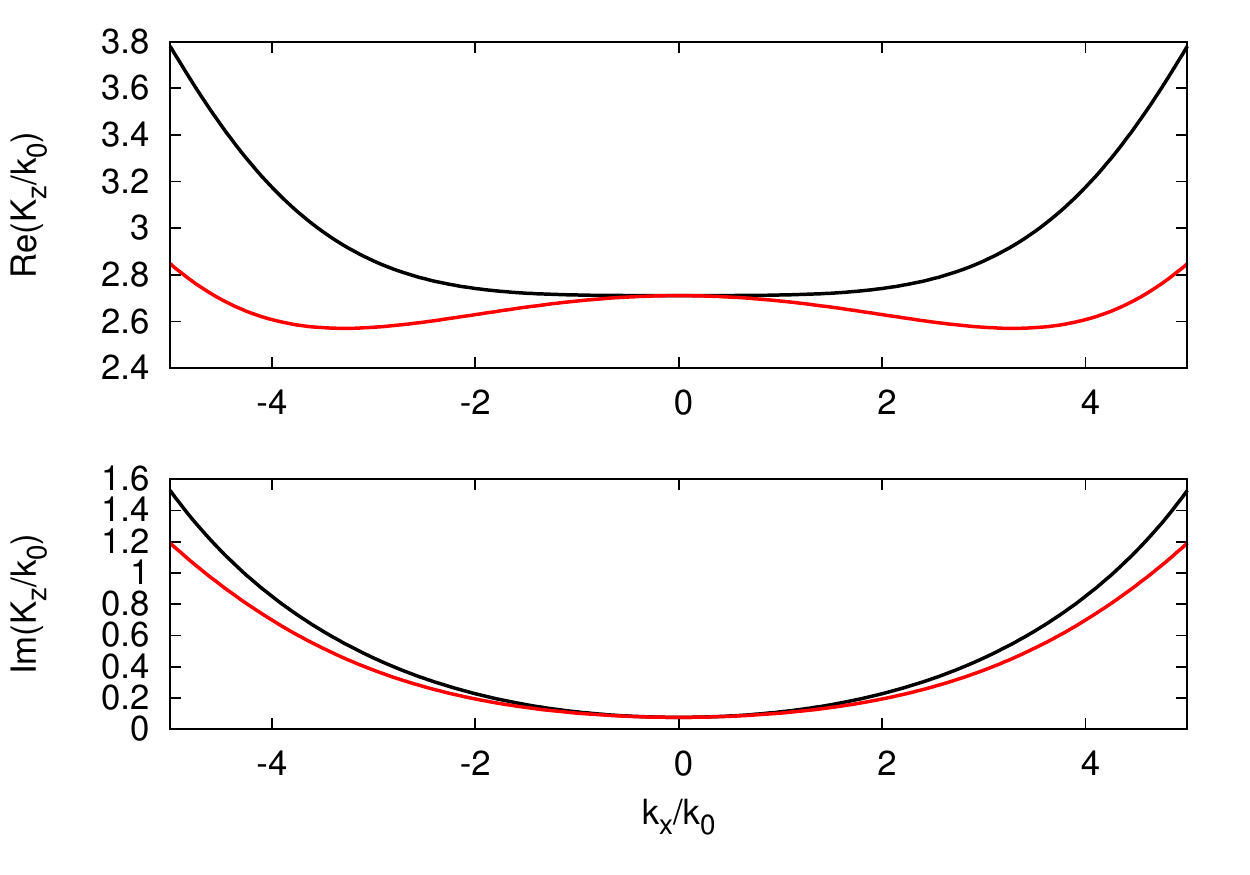}
\end{center}
\caption{Dispersion curves for a metallo-dielectric structure ($h_d=5$ nm, $h_m=22.6$ nm, $\lambda=363.8$ nm) in the canalization regime ({\em i.e.} a flat dispersion curve). Top: Real part. Bottom: Imaginary part. The local (black line) and nonlocal (red line) cases are shown. The difference is much more important here: nonlocality has a direct impact on the curvature of the dispersion curve.\label{fig:disp2}}
\end{figure}

Figure \ref{fig:disp2} shows that for such a structure, for which the isofrequency curve is expected to be flat or slightly negative, the nonlocal curve actually presents a clear positive curvature. We underline this means the structure, presenting a positive effective index, is unable to refocus light and cannot be used to build a lens at all.

Nonlocality has actually a double impact on the optical behavior of the structure: (i) the plasmonic effect (linked to the fact that the Poynting vector is opposite in the metal and in the dielectric\cite{benedicto11a}) is lowered by nonlocality, so that it generally takes thicker metallic layers to get the same effect and thus the same effective index and (ii) the longitudinal wave acts as a supplementary way for light to tunnel through metallic layers, so that the transmission is generally higher. The imaginary part of the Bloch wavevector, shown in Fig. \ref{fig:disp2}, is always smaller when the longitudinal wave is taken into account. This is not straightforward, since for gap-plasmons the plasmonic effect is lowered too, leading to a lower effective index, but the losses are higher for the same effective index\cite{moreau13}.

We underline that there is no real homogeneization regime when nonlocality is taken into account, because when the thickness of the metallic layers tends to zero, nonlocal effects intervene more and more. It should be much more difficult in that case to design a structure that would work exactly as desired since no simple formulae are available in that case to guide the design. Numerical optimization procedures are likely to be required. Another consequence is that using thinner layers does have an impact on the behavior of the structure, contrarily to the local situation, where it is equivalent in the homogeneization regime to take two different periods as long as the dielectric-to-metal ratio is kept constant. Even if this ratio may have to be raised a little bit to get the desired properties, at least the transmission is higher for thinner metallic layers due to nonlocality. From that point of view the effects of nonlocality can thus be considered as beneficial and speak in favor of the use of thinner metallic layers.

\subsection{Impact of nonlocality on a slab of -1 index metamaterial}

We have implemented the above scattering matrix algorithm, basing it on a code we have previously released\cite{krayzel10}, and that has previsouly been used to simulate the propagation of light beams in metallo-dielectric layers\cite{benedicto11b,benedicto12}. Here we consider the structure known to behave as a -1 index medium\cite{verhagen10} and that have been recently fabricated\cite{xu13}. This lens operation wavelength is of $363.8$ nm, in the close UV range. In that range the plasmonic effects are actually much higher\cite{benedicto11a}. They are usually linked to the Poynting vector inside the metal, that is roughly proportionnal to $\frac{1}{\epsilon}$, a factor that becomes important when the permittivity is negative but small. It is however not possible to consider much shorter wavelength because the titanium oxide then becomes much more absorbent. This is what makes this close UV range so interesting to build flat lenses - and this is a range for which the nonlocal effects are much more likely to be noticeable too.

\begin{figure}[h]
\begin{center}
\includegraphics[width=8cm]{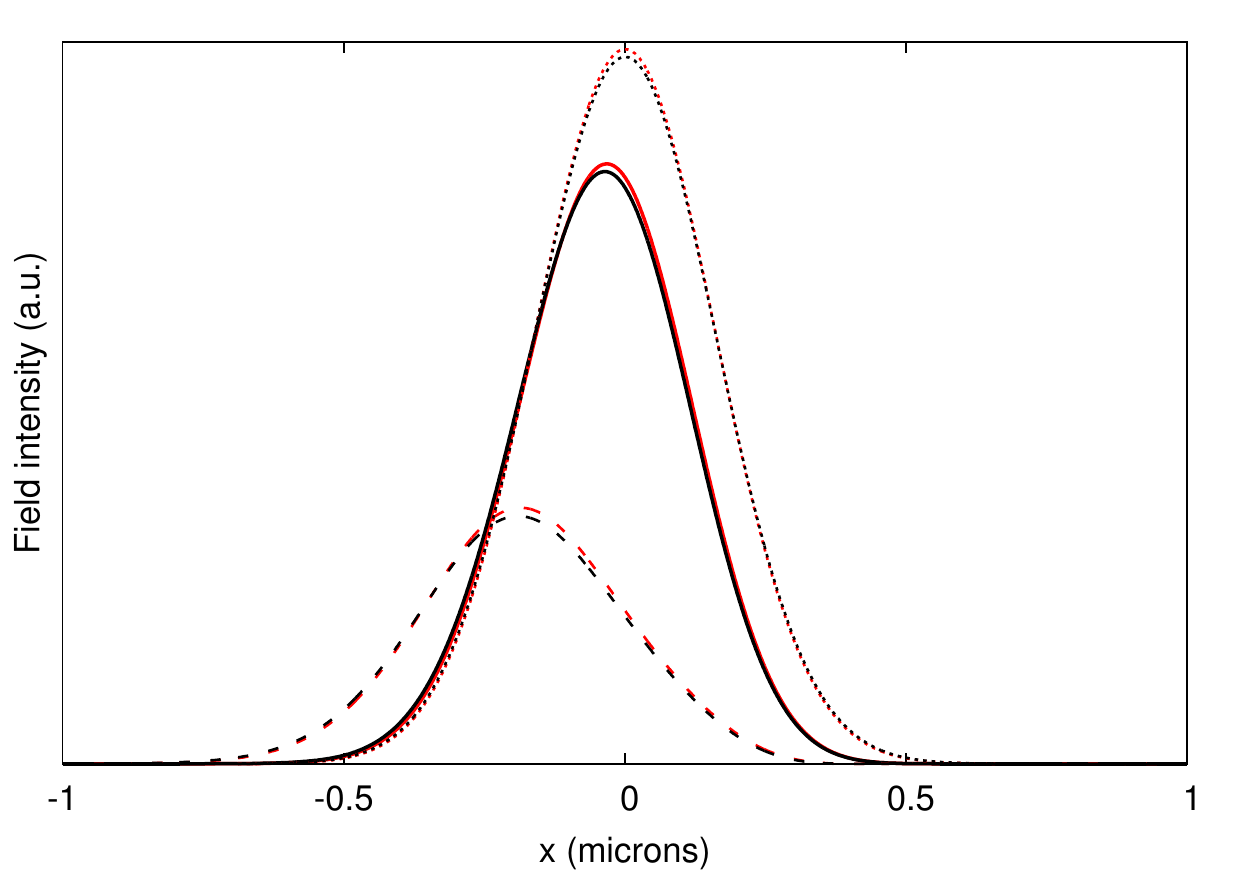}
\end{center}
\caption{Transmitted beam for different incidence angles (dotted line: normal incidence, solid line:$20^\circ$, dashed line: $40^\circ$) for the local (black) and the nonlocal (red) computation in the case of the -1 lens\cite{xu13}. The waist of the incident beam is equal to $1 \lambda$.\label{fig:angle}}
\end{figure}

\begin{figure}[h]
\begin{center}
\includegraphics[width=8cm]{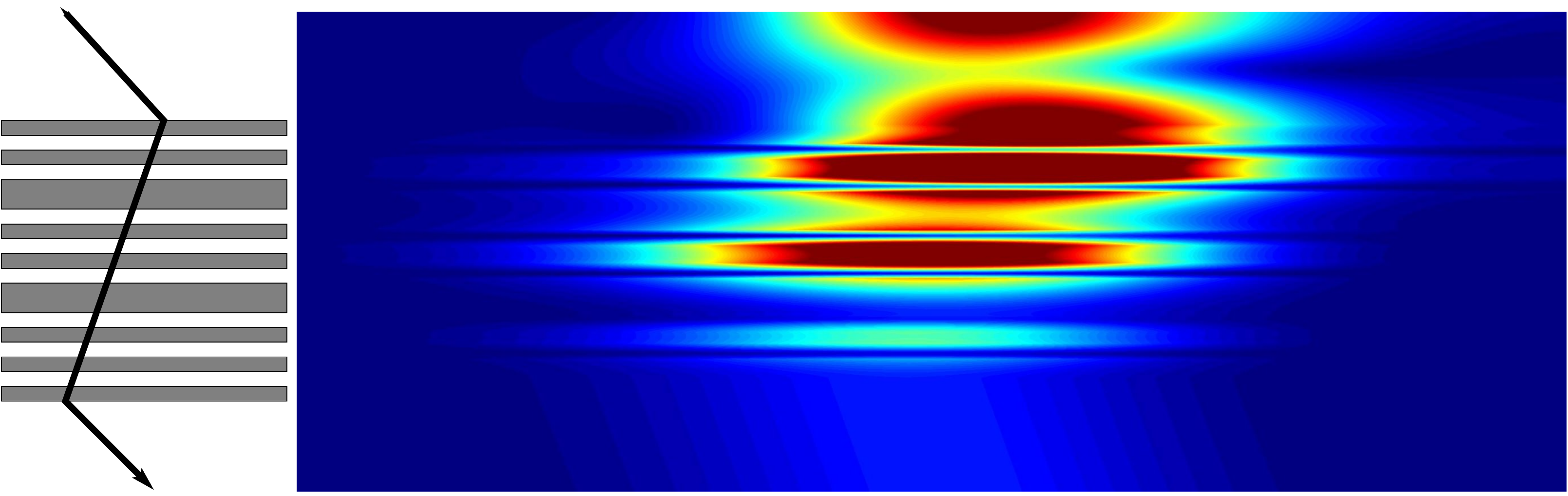}
\end{center}
\caption{Propagation of an incident Gaussian beam with a waist of $1\lambda$ through a -1 index lens\cite{xu13}, showing the
negative refraction phenomenon, illustrated on the left. The modulus of the magnetic field is plotted on the right.\label{fig:joli}}
\end{figure}

We have simulated what happens when the structure is illuminated with a Gaussian beam with a non-normal incidence, an experiment that has been made to study the negative refraction. A typical result is shown figure \ref{fig:joli}. In order to compare local and nonlocal simulations, we have plotted figure \ref{fig:angle} different beam profiles for different incidence angles at the exit of the lens. There is actually very little difference between the two. This can be related to the fact that (i) we have seen above that for hyperbolic media with a -1 effective index nonlocality has not a huge effect on the dispersion relation and (ii) nonlocal effects can be seen for large wavevectors that are not concerned by this kind of experiment.

\subsection{Impact in the canalization regime}

\begin{figure}[h]
\begin{center}
\includegraphics[width=8cm]{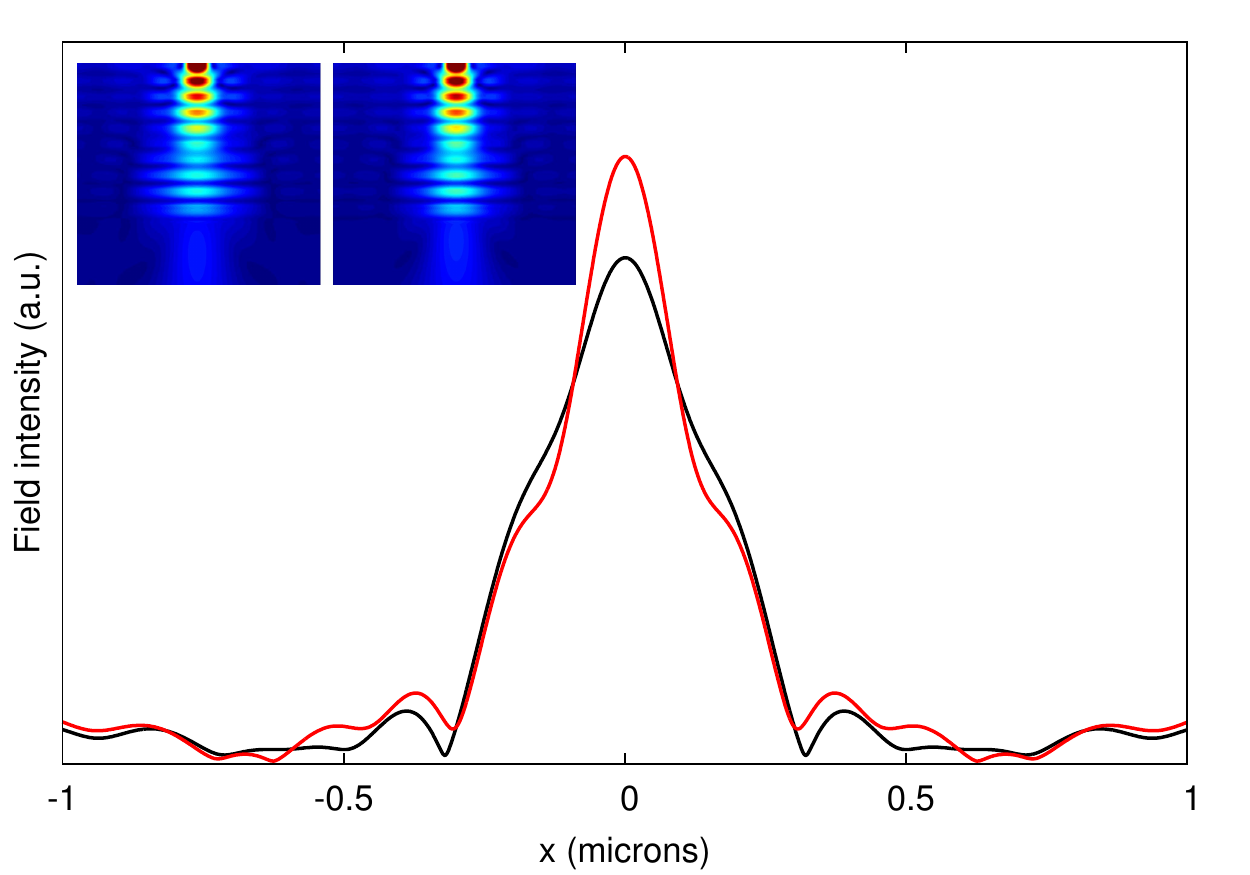}
\end{center}
\caption{Profile of the outgoing beam (magnetic field) for the local (black) and nonlocal (red) computation (case (i)). The field profile is computed at the very edge of the lens. Inset: local (left) and nonlocal (right) corresponding field maps for the magnetic field.
\label{fig:profil1}}
\end{figure}

\begin{figure}[h]
\begin{center}
\includegraphics[width=8cm]{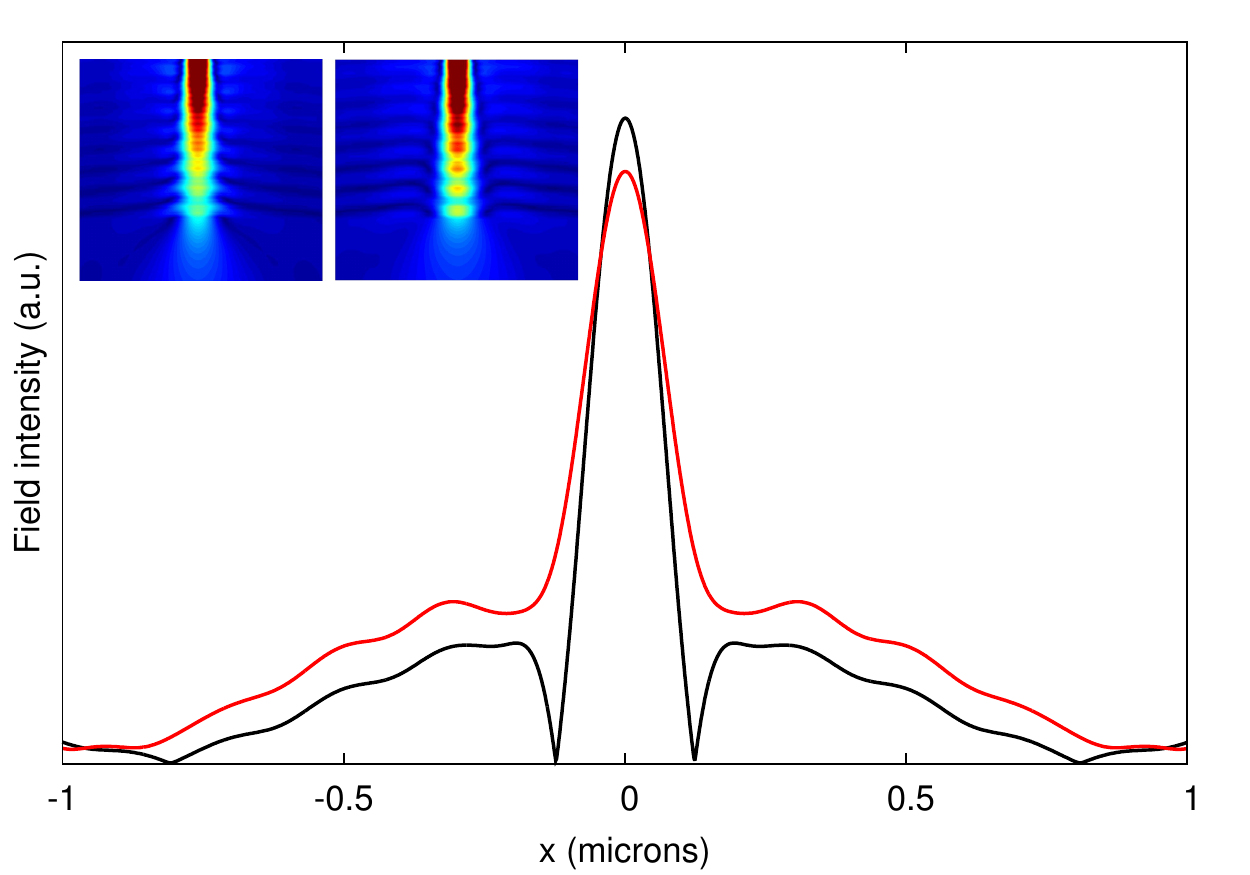}
\end{center}
\caption{Profile of the outgoing beam for the local (black) and nonlocal (red) computation (case (ii)). The field profile is computed at the very edge of the lens. Inset: local (left) and nonlocal (right) corresponding field maps for the magnetic field. 
\label{fig:profil2}}
\end{figure}

Keeping the same materials and operation wavelength, it is possible to find a structure in the canalization regime, {\em i.e.} presenting a flat dispersion curve according to the local theory. In such a medium, all the waves (whether they are evanescent in the outside medium or not) propagate, and they do it in the same direction. As long as the ratio $\frac{h_d}{h_m}$ is equal to the ratio $\left|\frac{\epsilon_d}{\epsilon_m}\right|\simeq 4.2$, and the overall period stays small with respect to the wavelength, we can consider that we are in the canalization regime. Two cases are considered here (i) a 10 period structure with 10 nm metallic layer and a 42 nm dielectric layer and (ii) a 25 period structure with a 4 nm metallic layer and a 16.8 nm thick dielectric layer. Both structures begin and terminate with a metallic layer so that their respective thicknesses are 530 nm and 524 nm. The structure is illuminated with a Gaussian beam (normal incidence, wavelength of $\lambda=363.8$ nm, waist of $0.1$ nm and focused on the entrance of the structure). It should be underlined that what we call a Gaussian beam contains evanescents, so that it is actually almost a ponctual source. The results of the computation are shown Fig. \ref{fig:profil1} for the first case with 10 nm thick metallic layers and Fig. \ref{fig:profil2} for the second case with thinner layers.

The analysis of the dispersion curves is confirmed: the thinner the metallic layers, the more light manages to go through. Another point is that the predictions of the local theory are significantly different from the nonlocal one when it comes to the profile of the outgoing beam. This means that nonlocality can definitely not be ignored when the whole purpose of the structure is to make an image of a source with subwavelength resolution.  

\subsection{Gap-Plasmon Resonance}

In many situations, nonlocality does not have a noticeable impact on the optical properties of a structure. Using the same structures as considered above, but for larger wavelength (outside of the close UV range mentioned above) makes that impact very often negligible.

 In the eighties, the community was hoping that surface plasmons would be sensitive to nonlocality\cite{forstmann86}, but this is not the case because the wavevector of the surface plasmon cannot reach high enough values. The gap-plasmon is the fundamental mode of a metallic waveguide\cite{bozhevolnyi07} and it experiences a violent plasmonic slowdown when the size of the gap is decreased. Its wavevector can theoretically become arbitrarily large for a small enough gap. This makes this mode very sensitive to nonlocality\cite{moreau13} when the wavelength approaches the electrons mean free path\cite{chapuis08}. A way to excite the gap-plasmon is to use gap-plasmon resonators\cite{moreau12b,moreau13}, but a more conventional way would be to couple the gap-plasmon using a prism. It is more difficult to reach extremely high wavevectors with such a setup. However this drawback would be compensated by a very high control of the experimental conditions so that the nonlocal effect should easily be measurable.

\begin{figure}[h]
\begin{center}
\includegraphics[width=8cm]{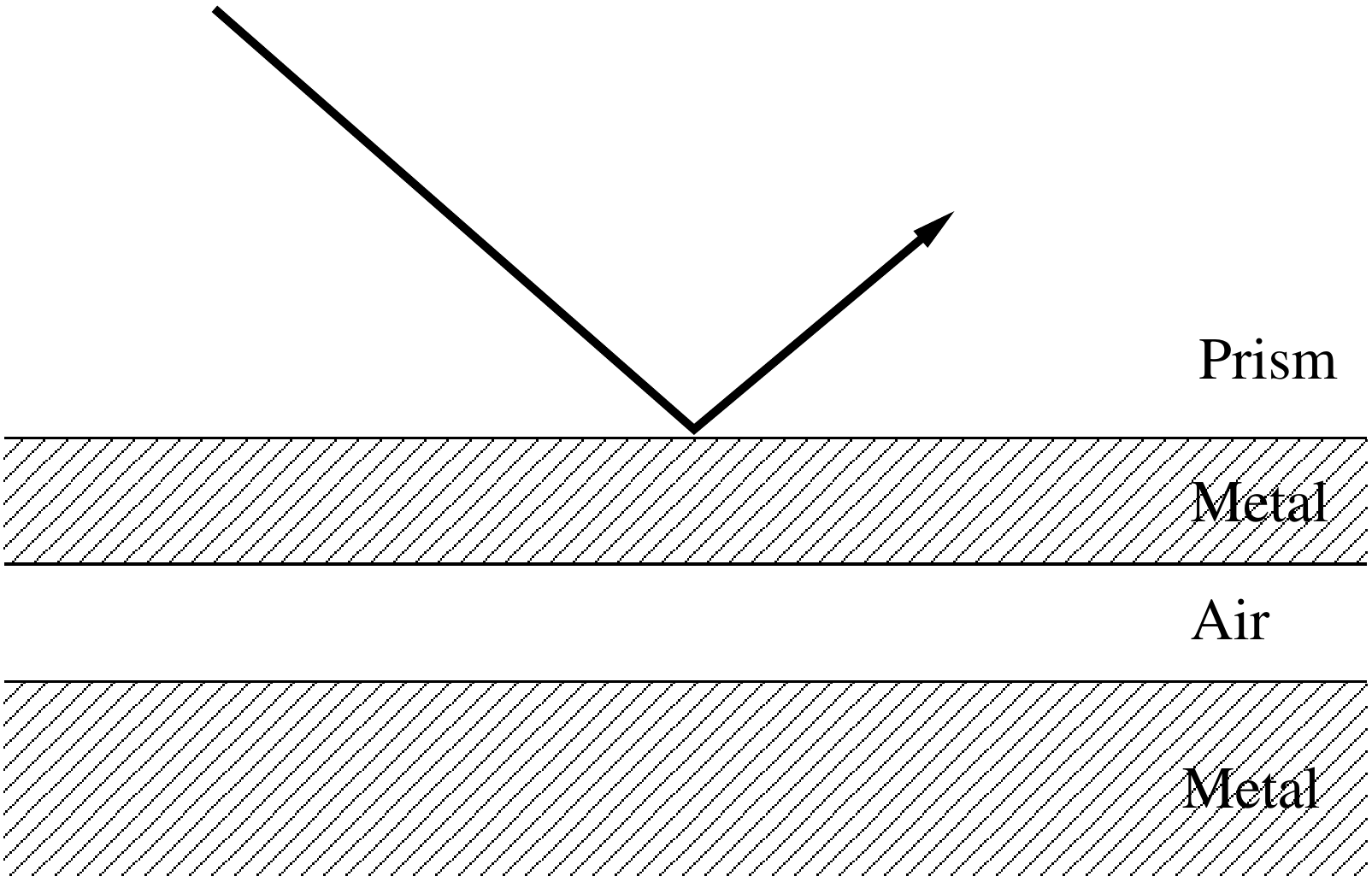}
\end{center}
\caption{Diagram of the excitation of a gap-plasmon using a prism coupler. The thickness of the upper metallic layer is 18 nm. 
\label{fig:schema}}
\end{figure}

Here we consider the case of a gap-plasmon excited using a $TiO_2$ prism\cite{devore51}. The prism is placed above, on top of a 18 nm thick silver layer, see Fig. \ref{fig:schema}. Two gap sizes are first considered here (10 and 12 nm respectively).

\begin{figure}[h]
\begin{center}
\includegraphics[width=8cm]{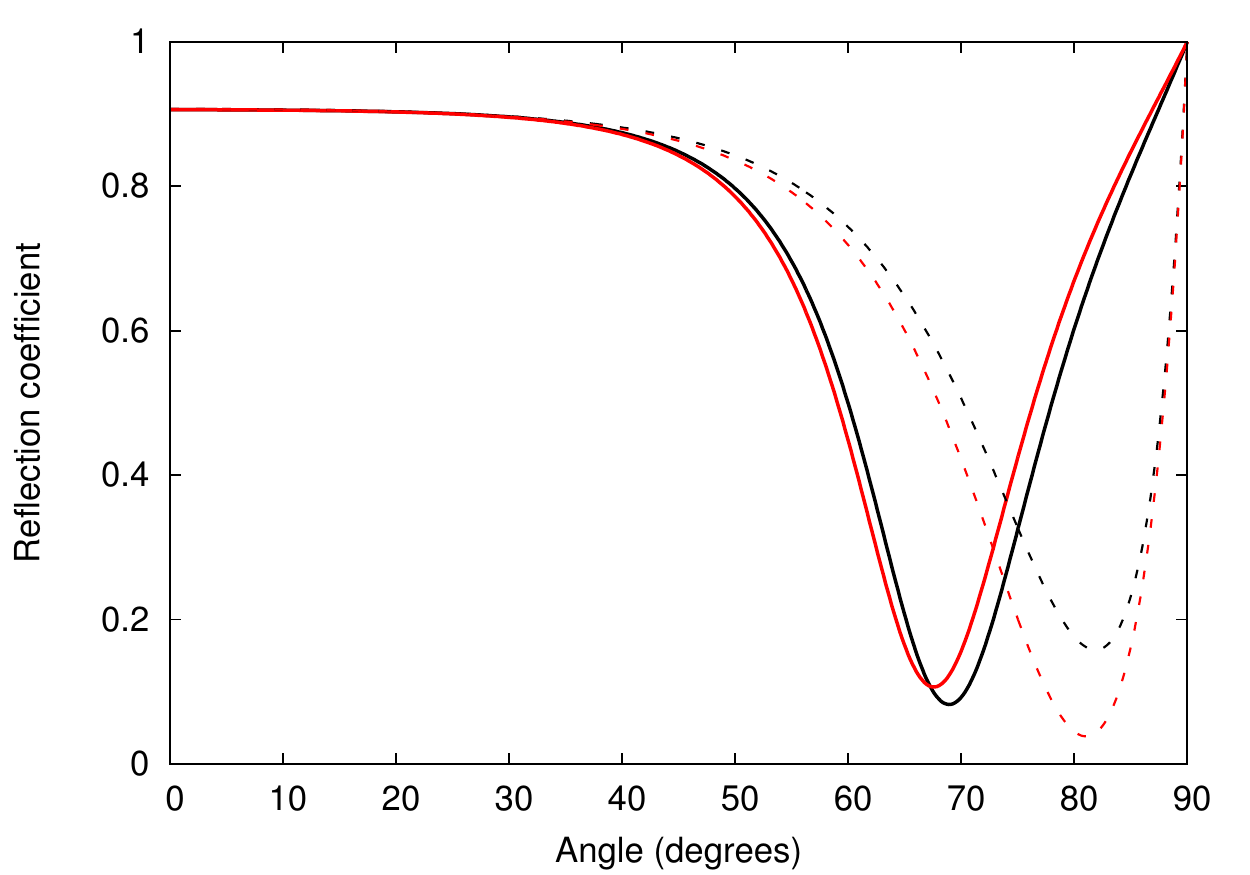}
\end{center}
\caption{Reflection coefficient of the structure described in Fig. \ref{fig:schema} for a 12 nm gap (solid lines) in the local (black) and nonlocal (red) case and for a 10 nm (dashed lines) in the local (black) and nonlocal (red) case.
\label{fig:prismr}}
\end{figure}

\begin{figure}[h]
\begin{center}
\includegraphics[width=8cm]{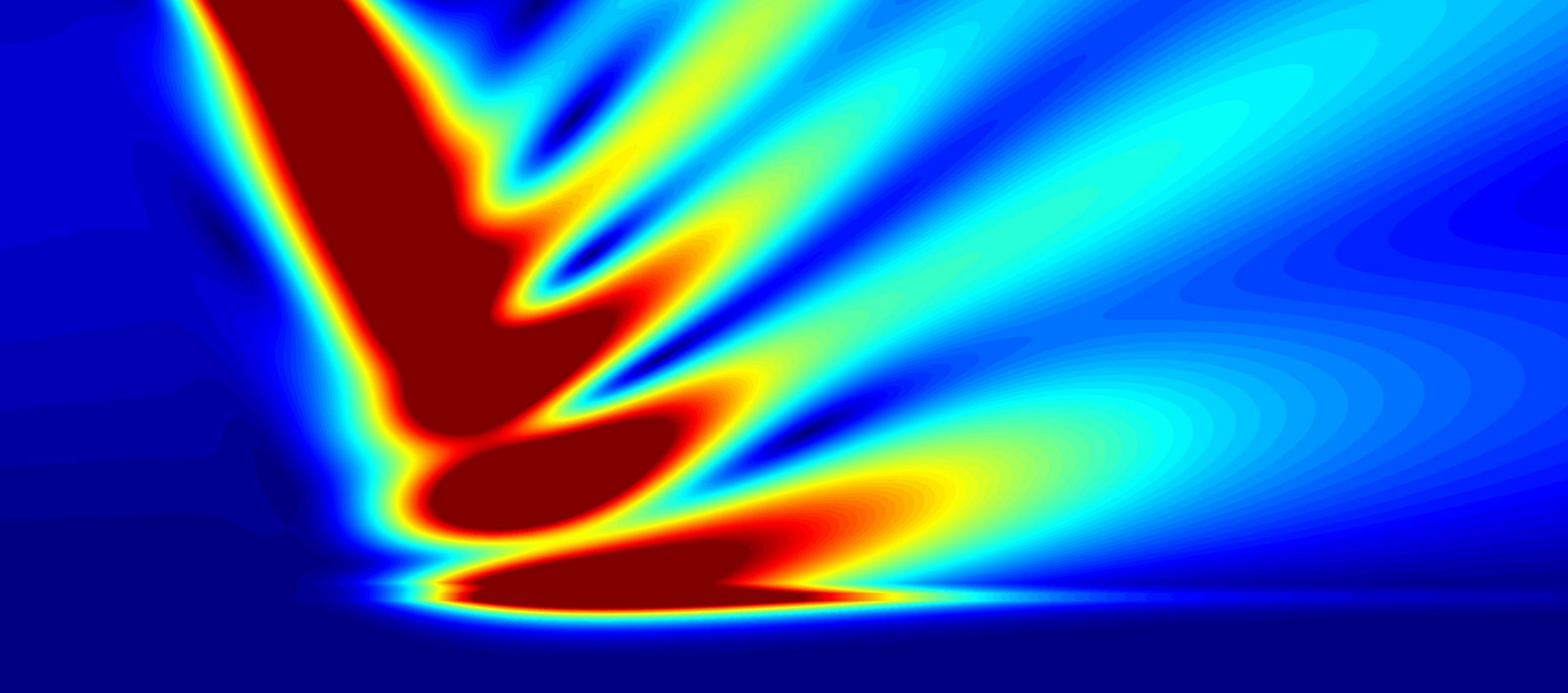}
\end{center}
\caption{Modulus of the $H$ field when a GPR is excited, in the nonlocal case, for a $67.6^\circ$  incidence angle and a 12 nm gap.
\label{fig:prism}}
\end{figure}

The structure is illuminated by a Gaussian beam (wavelength of $\lambda=543$ nm, waist of $\lambda$ and focused on the entrance of the structure) and the incidence angle is made to vary. 
The results are shown in Fig. \ref{fig:prismr}. The gap-plasmon resonance (GPR) can clearly be seen for high incidence angle, and nonlocality has obviously a strong impact on the reflection coefficient of the structure. For a 12 nm gap, the main difference is a shift in the resonance : the GPR is excited for a smaller angle when nonlocality is taken into account ($67.6^\circ$ instead of $68.9^\circ$). The corresponding field is shown in Fig. \ref{fig:prism} For a 10 nm gap, the GPR is excited for very high incidence angles, so that the main difference there is more the depth of the resonance than its position. 

\begin{figure}[h]
\begin{center}
\includegraphics[width=8cm]{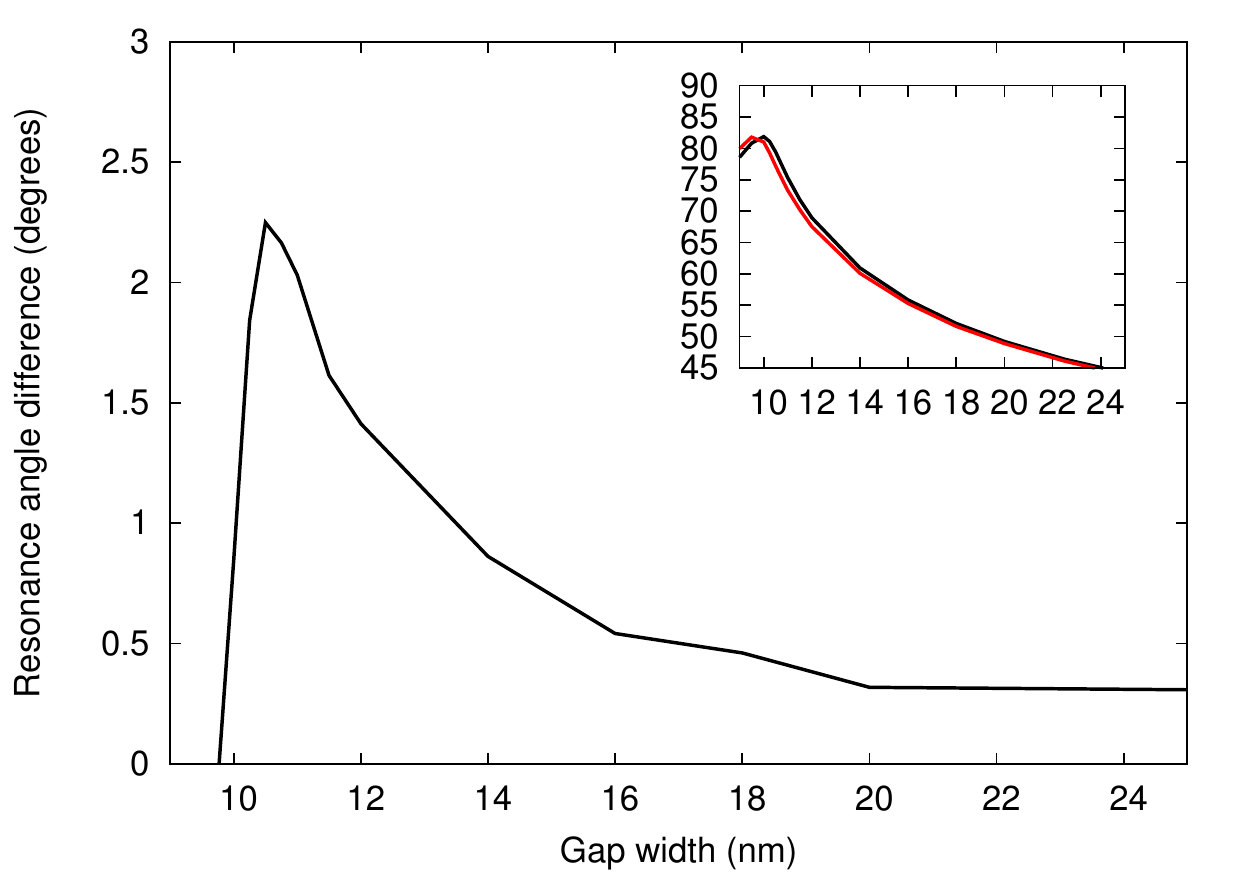}
\end{center}
\caption{Difference in degrees between the local predictions and the nonlocal ones, regarding the exact position of the gap-plasmon resonance. Inset: Angular position (in degrees) of the gap-plasmon resonance (GPR) as a function of the gap width (in nm).\label{fig:gpr}}
\end{figure}

Figure \ref{fig:gpr} shows the angle for which the GPR can be excited as a function of the gap width, for a $TiO_2$ prism covered with a 18 nm thick gold layer. For larger gaps, the difference between the local and the nonlocal theory can hardly be distinguished, but for gaps that are close to 10 nm the difference is quite obvious. Below 9 nm, the gap-plasmon has actually too high an effective index to be excited using a prism.

The difference between the local predictions and the nonlocal ones, however small, should be totally measurable.

We underline here that, for gaps that are of the order of 10 nm, any influence of the tunneling effect, and probably of the spill-out can totally be excluded. The effect relies purely on the fact that the gap-plasmon resonance is slowed down by plasmonic effects so that the mode becomes sensitive to nonlocality. This kind of very simple experiment may bring definitive answers regarding the validity of the different descriptions of nonlocality that are currently available. 

\section{Conclusion}

We have presented here numerical tools that allow to take nonlocality in metals into account when simulating the propagation of a plane wave or of a beam in a metallo-dielectric multilayer. These tools, relying essentially on analytical calculations, are meant to be as accurate as possible - through the use of accurate material characteristics\cite{rakic98} and boundary conditions that can be considered conservative\cite{moreau13} compared to other implementation of the hydrodynamic model. The formula presented here are easy to adapt for different descriptions of the metal and different boundary conditions so that previous results can be retrieved and checked using the present work. Furthermore, if the hydrodynamic model needs further tuning to match future experiments, the present formalism should be very easy to adapt. Finally, we have made the codes we have written freely available\cite{pgp}.

We have used these tools to assess the impact of nonlocality on realistic metallo-dielectric structures presenting a negative refractive index and in the canalization regime. Our conclusion are that for a negative index around $-1$ the impact of nonlocality should be expected to be negligible. For higher absolute values of the refractive index that are required to reach subwalength resolution, and especially in the canalization regime, the effect of nonlocality cannot be ignored. We underline that even small effects like the small change in the effective index due to nonlocality will have an impact on the operation of flat lenses, especially when they are able to reach super-resolution\cite{benedicto12}. In that case, the propagation of high wavevector waves are actually responsible for the subwavelength resolution\cite{belov05,belov06}. The structure has thus to be finely optimized\cite{benedicto13} and there is little doubt that nonlocal effects should be taken into account. The tools we have provided here should help to finely simulate the optical behaviour of such structures. 

Finally, we have shown that nonlocality has a small but definitely measurable impact on the excitation of a gap-plasmon using evancescent coupling. Such an experiment could bring clear and definitive answers to many questions regarding nonlocality, including the value of the $\beta$ parameter or the right boundary condition. The gaps that are considered here are very large compared to previously studied situations\cite{toscano12,ciraci12,savage12,teperik13} so that the measured effects can be attributed solely to nonlocality. At these scales, tunneling can be excluded, and the spill-out is not likely to have any noticeable impact\cite{toscano14}.

We hope that the present work will make it easy for the community to assess the impact of nonlocality thoroughly and to take it accurately into account, in order to design structures or experiments based on metallo-dielectric multilayers.

\section*{Acknowledgments}

This work has been supported by the French National Research Agency, ``Physics of Gap-Plasmons'' project number ANR-13-JS10-0003.

\bibliography{nonlocal}

\begin{thebibliography}{10}%
\makeatletter
\providecommand \@ifxundefined [1]{%
 \ifx #1\undefined \expandafter \@firstoftwo
 \else \expandafter \@secondoftwo
\fi
}%
\providecommand \@ifnum [1]{%
 \ifnum #1\expandafter \@firstoftwo
 \else \expandafter \@secondoftwo
\fi
}%
\providecommand \enquote [1]{``#1''}%
\providecommand \bibnamefont  [1]{#1}%
\providecommand \bibfnamefont [1]{#1}%
\providecommand \citenamefont [1]{#1}%
\providecommand\href[0]{\@sanitize\@href}%
\providecommand\@href[1]{\endgroup\@@startlink{#1}\endgroup\@@href}%
\providecommand\@@href[1]{#1\@@endlink}%
\providecommand \@sanitize [0]{\begingroup\catcode`\&12\catcode`\#12\relax}%
\@ifxundefined \pdfoutput {\@firstoftwo}{%
 \@ifnum{\z@=\pdfoutput}{\@firstoftwo}{\@secondoftwo}%
}{%
 \providecommand\@@startlink[1]{\leavevmode\special{html:<a href="#1">}}%
 \providecommand\@@endlink[0]{\special{html:</a>}}%
}{%
 \providecommand\@@startlink[1]{%
  \leavevmode
  \pdfstartlink
   attr{/Border[0 0 1 ]/H/I/C[0 1 1]}%
   user{/Subtype/Link/A<</Type/Action/S/URI/URI(#1)>>}%
  \relax
 }%
 \providecommand\@@endlink[0]{\pdfendlink}%
}%
\providecommand \url  [0]{\begingroup\@sanitize \@url }%
\providecommand \@url [1]{\endgroup\@href {#1}{\urlprefix}}%
\providecommand \urlprefix [0]{URL }%
\providecommand \Eprint[0]{\href }%
\@ifxundefined \urlstyle {%
  \providecommand \doi [1]{doi:\discretionary{}{}{}#1}%
}{%
  \providecommand \doi [0]{doi:\discretionary{}{}{}\begingroup
  \urlstyle{rm}\Url }%
}%
\providecommand \doibase [0]{http://dx.doi.org/}%
\providecommand \Doi[1]{\href{\doibase#1}}%
\providecommand \bibAnnote [3]{%
  \BibitemShut{#1}%
  \begin{quotation}\noindent
    \textsc{Key:}\ #2\\\textsc{Annotation:}\ #3%
  \end{quotation}%
}%
\providecommand \bibAnnoteFile [2]{%
  \IfFileExists{#2}{\bibAnnote {#1} {#2} {\input{#2}}}{}%
}%
\providecommand \typeout [0]{\immediate \write \m@ne }%
\providecommand \selectlanguage [0]{\@gobble}%
\providecommand \bibinfo [0]{\@secondoftwo}%
\providecommand \bibfield [0]{\@secondoftwo}%
\providecommand \translation [1]{[#1]}%
\providecommand \BibitemOpen[0]{}%
\providecommand \bibitemStop [0]{}%
\providecommand \bibitemNoStop [0]{.\EOS\space}%
\providecommand \EOS [0]{\spacefactor3000\relax}%
\providecommand \BibitemShut [1]{\csname bibitem#1\endcsname}%
\bibitem{scalora98}%
  \BibitemOpen
  \bibfield{author}{%
  \bibinfo {author} {\bibfnamefont{M.}~\bibnamefont{Scalora}}, \bibinfo
  {author} {\bibfnamefont{M.}~\bibnamefont{Bloemer}}, \bibinfo {author}
  {\bibfnamefont{A.}~\bibnamefont{Pethel}}, \bibinfo {author}
  {\bibfnamefont{J.}~\bibnamefont{Dowling}}, \bibinfo {author}
  {\bibfnamefont{C.}~\bibnamefont{Bowden}},\ and\ \bibinfo {author}
  {\bibfnamefont{A.}~\bibnamefont{Manka}},\ }%
  \bibfield{journal}{%
  \bibinfo {journal} {Journal of Applied Physics}\ }%
  \textbf{\bibinfo {volume} {83}},\ \bibinfo {pages} {2377} (\bibinfo {year}
  {1998})%
  \bibAnnoteFile{NoStop}{scalora98}%
\bibitem{cai2005superlens}%
  \BibitemOpen
  \bibfield{author}{%
  \bibinfo {author} {\bibfnamefont{W.}~\bibnamefont{Cai}}, \bibinfo {author}
  {\bibfnamefont{D.~A.}\ \bibnamefont{Genov}},\ and\ \bibinfo {author}
  {\bibfnamefont{V.~M.}\ \bibnamefont{Shalaev}},\ }%
  \bibfield{journal}{%
  \bibinfo {journal} {Physical review B}\ }%
  \textbf{\bibinfo {volume} {72}},\ \bibinfo {pages} {193101} (\bibinfo {year}
  {2005})%
  \bibAnnoteFile{NoStop}{cai2005superlens}%
\bibitem{scalora07}%
  \BibitemOpen
  \bibfield{author}{%
  \bibinfo {author} {\bibfnamefont{M.}~\bibnamefont{Scalora}}, \bibinfo
  {author} {\bibfnamefont{G.}~\bibnamefont{D'Aguanno}}, \bibinfo {author}
  {\bibfnamefont{N.}~\bibnamefont{Mattiucci}}, \bibinfo {author}
  {\bibfnamefont{M.~J.}\ \bibnamefont{Bloemer}}, \bibinfo {author}
  {\bibfnamefont{D.}~\bibnamefont{de~Ceglia}}, \bibinfo {author}
  {\bibfnamefont{M.}~\bibnamefont{Centini}}, \bibinfo {author}
  {\bibfnamefont{A.}~\bibnamefont{Mandatori}}, \bibinfo {author}
  {\bibfnamefont{C.}~\bibnamefont{Sibilia}}, \bibinfo {author}
  {\bibfnamefont{N.}~\bibnamefont{Akozbek}}, \bibinfo {author}
  {\bibfnamefont{M.~G.}\ \bibnamefont{Cappeddu}}, \emph{et~al.},\ }%
  \bibfield{journal}{%
  \bibinfo {journal} {Optics Express}\ }%
  \textbf{\bibinfo {volume} {15}},\ \bibinfo {pages} {508} (\bibinfo {year}
  {2007})%
  \bibAnnoteFile{NoStop}{scalora07}%
\bibitem{hoffman2007negative}%
  \BibitemOpen
  \bibfield{author}{%
  \bibinfo {author} {\bibfnamefont{A.~J.}\ \bibnamefont{Hoffman}}, \bibinfo
  {author} {\bibfnamefont{L.}~\bibnamefont{Alekseyev}}, \bibinfo {author}
  {\bibfnamefont{S.~S.}\ \bibnamefont{Howard}}, \bibinfo {author}
  {\bibfnamefont{K.~J.}\ \bibnamefont{Franz}}, \bibinfo {author}
  {\bibfnamefont{D.}~\bibnamefont{Wasserman}}, \bibinfo {author}
  {\bibfnamefont{V.~A.}\ \bibnamefont{Podolskiy}}, \bibinfo {author}
  {\bibfnamefont{E.~E.}\ \bibnamefont{Narimanov}}, \bibinfo {author}
  {\bibfnamefont{D.~L.}\ \bibnamefont{Sivco}},\ and\ \bibinfo {author}
  {\bibfnamefont{C.}~\bibnamefont{Gmachl}},\ }%
  \bibfield{journal}{%
  \bibinfo {journal} {Nature materials}\ }%
  \textbf{\bibinfo {volume} {6}},\ \bibinfo {pages} {946} (\bibinfo {year}
  {2007})%
  \bibAnnoteFile{NoStop}{hoffman2007negative}%
\bibitem{liu2007far}%
  \BibitemOpen
  \bibfield{author}{%
  \bibinfo {author} {\bibfnamefont{Z.}~\bibnamefont{Liu}}, \bibinfo {author}
  {\bibfnamefont{H.}~\bibnamefont{Lee}}, \bibinfo {author}
  {\bibfnamefont{Y.}~\bibnamefont{Xiong}}, \bibinfo {author}
  {\bibfnamefont{C.}~\bibnamefont{Sun}},\ and\ \bibinfo {author}
  {\bibfnamefont{X.}~\bibnamefont{Zhang}},\ }%
  \bibfield{journal}{%
  \bibinfo {journal} {science}\ }%
  \textbf{\bibinfo {volume} {315}},\ \bibinfo {pages} {1686} (\bibinfo {year}
  {2007})%
  \bibAnnoteFile{NoStop}{liu2007far}%
\bibitem{noginov2009bulk}%
  \BibitemOpen
  \bibfield{author}{%
  \bibinfo {author} {\bibfnamefont{M.}~\bibnamefont{Noginov}}, \bibinfo
  {author} {\bibfnamefont{Y.~A.}\ \bibnamefont{Barnakov}}, \bibinfo {author}
  {\bibfnamefont{G.}~\bibnamefont{Zhu}}, \bibinfo {author}
  {\bibfnamefont{T.}~\bibnamefont{Tumkur}}, \bibinfo {author}
  {\bibfnamefont{H.}~\bibnamefont{Li}},\ and\ \bibinfo {author}
  {\bibfnamefont{E.}~\bibnamefont{Narimanov}},\ }%
  \bibfield{journal}{%
  \bibinfo {journal} {Applied Physics Letters}\ }%
  \textbf{\bibinfo {volume} {94}},\ \bibinfo {pages} {151105} (\bibinfo {year}
  {2009})%
  \bibAnnoteFile{NoStop}{noginov2009bulk}%
\bibitem{cortes12}%
  \BibitemOpen
  \bibfield{author}{%
  \bibinfo {author} {\bibfnamefont{C.}~\bibnamefont{Cortes}}, \bibinfo {author}
  {\bibfnamefont{W.}~\bibnamefont{Newman}}, \bibinfo {author}
  {\bibfnamefont{S.}~\bibnamefont{Molesky}},\ and\ \bibinfo {author}
  {\bibfnamefont{Z.}~\bibnamefont{Jacob}},\ }%
  \bibfield{journal}{%
  \bibinfo {journal} {Journal of Optics}\ }%
  \textbf{\bibinfo {volume} {14}},\ \bibinfo {pages} {063001} (\bibinfo {year}
  {2012})%
  \bibAnnoteFile{NoStop}{cortes12}%
\bibitem{xu13}%
  \BibitemOpen
  \bibfield{author}{%
  \bibinfo {author} {\bibfnamefont{T.}~\bibnamefont{Xu}}, \bibinfo {author}
  {\bibfnamefont{A.}~\bibnamefont{Agrawal}}, \bibinfo {author}
  {\bibfnamefont{M.}~\bibnamefont{Abashin}}, \bibinfo {author}
  {\bibfnamefont{K.~J.}\ \bibnamefont{Chau}},\ and\ \bibinfo {author}
  {\bibfnamefont{H.~J.}\ \bibnamefont{Lezec}},\ }%
  \bibfield{journal}{%
  \bibinfo {journal} {Nature}\ }%
  \textbf{\bibinfo {volume} {497}},\ \bibinfo {pages} {470} (\bibinfo {year}
  {2013})%
  \bibAnnoteFile{NoStop}{xu13}%
\bibitem{biehs12}%
  \BibitemOpen
  \bibfield{author}{%
  \bibinfo {author} {\bibfnamefont{S.-A.}\ \bibnamefont{Biehs}}, \bibinfo
  {author} {\bibfnamefont{M.}~\bibnamefont{Tschikin}},\ and\ \bibinfo {author}
  {\bibfnamefont{P.}~\bibnamefont{Ben-Abdallah}},\ }%
  \bibfield{journal}{%
  \bibinfo {journal} {Physical review letters}\ }%
  \textbf{\bibinfo {volume} {109}},\ \bibinfo {pages} {104301} (\bibinfo {year}
  {2012})%
  \bibAnnoteFile{NoStop}{biehs12}%
\bibitem{guo12}%
  \BibitemOpen
  \bibfield{author}{%
  \bibinfo {author} {\bibfnamefont{Y.}~\bibnamefont{Guo}}, \bibinfo {author}
  {\bibfnamefont{C.~L.}\ \bibnamefont{Cortes}}, \bibinfo {author}
  {\bibfnamefont{S.}~\bibnamefont{Molesky}},\ and\ \bibinfo {author}
  {\bibfnamefont{Z.}~\bibnamefont{Jacob}},\ }%
  \bibfield{journal}{%
  \bibinfo {journal} {Applied Physics Letters}\ }%
  \textbf{\bibinfo {volume} {101}},\ \bibinfo {pages} {131106} (\bibinfo {year}
  {2012})%
  \bibAnnoteFile{NoStop}{guo12}%
\bibitem{narimanov2012beyond}%
  \BibitemOpen
  \bibfield{author}{%
  \bibinfo {author} {\bibfnamefont{E.}~\bibnamefont{Narimanov}}\ and\ \bibinfo
  {author} {\bibfnamefont{I.}~\bibnamefont{Smolyaninov}},\ }%
  in\ \emph{\bibinfo {booktitle} {Quantum Electronics and Laser Science
  Conference}}\ (\bibinfo {organization} {Optical Society of America},\
  \bibinfo {year} {2012})\ pp.\ \bibinfo {pages} {QM2E--1}%
  \bibAnnoteFile{NoStop}{narimanov2012beyond}%
\bibitem{fang2005sub}%
  \BibitemOpen
  \bibfield{author}{%
  \bibinfo {author} {\bibfnamefont{N.}~\bibnamefont{Fang}}, \bibinfo {author}
  {\bibfnamefont{H.}~\bibnamefont{Lee}}, \bibinfo {author}
  {\bibfnamefont{C.}~\bibnamefont{Sun}},\ and\ \bibinfo {author}
  {\bibfnamefont{X.}~\bibnamefont{Zhang}},\ }%
  \bibfield{journal}{%
  \bibinfo {journal} {Science}\ }%
  \textbf{\bibinfo {volume} {308}},\ \bibinfo {pages} {534} (\bibinfo {year}
  {2005})%
  \bibAnnoteFile{NoStop}{fang2005sub}%
\bibitem{belov05}%
  \BibitemOpen
  \bibfield{author}{%
  \bibinfo {author} {\bibfnamefont{P.~A.}\ \bibnamefont{Belov}}, \bibinfo
  {author} {\bibfnamefont{C.~R.}\ \bibnamefont{Simovski}},\ and\ \bibinfo
  {author} {\bibfnamefont{P.}~\bibnamefont{Ikonen}},\ }%
  \bibfield{journal}{%
  \bibinfo {journal} {Physical review B}\ }%
  \textbf{\bibinfo {volume} {71}},\ \bibinfo {pages} {193105} (\bibinfo {year}
  {2005})%
  \bibAnnoteFile{NoStop}{belov05}%
\bibitem{jacob06}%
  \BibitemOpen
  \bibfield{author}{%
  \bibinfo {author} {\bibfnamefont{Z.}~\bibnamefont{Jacob}}, \bibinfo {author}
  {\bibfnamefont{L.~V.}\ \bibnamefont{Alekseyev}},\ and\ \bibinfo {author}
  {\bibfnamefont{E.}~\bibnamefont{Narimanov}},\ }%
  \bibfield{journal}{%
  \bibinfo {journal} {Optics express}\ }%
  \textbf{\bibinfo {volume} {14}},\ \bibinfo {pages} {8247} (\bibinfo {year}
  {2006})%
  \bibAnnoteFile{NoStop}{jacob06}%
\bibitem{belov06}%
  \BibitemOpen
  \bibfield{author}{%
  \bibinfo {author} {\bibfnamefont{P.~A.}\ \bibnamefont{Belov}}\ and\ \bibinfo
  {author} {\bibfnamefont{Y.}~\bibnamefont{Hao}},\ }%
  \bibfield{journal}{%
  \bibinfo {journal} {Physical Review B}\ }%
  \textbf{\bibinfo {volume} {73}},\ \bibinfo {pages} {113110} (\bibinfo {year}
  {2006})%
  \bibAnnoteFile{NoStop}{belov06}%
\bibitem{smolyaninov2007magnifying}%
  \BibitemOpen
  \bibfield{author}{%
  \bibinfo {author} {\bibfnamefont{I.~I.}\ \bibnamefont{Smolyaninov}}, \bibinfo
  {author} {\bibfnamefont{Y.-J.}\ \bibnamefont{Hung}},\ and\ \bibinfo {author}
  {\bibfnamefont{C.~C.}\ \bibnamefont{Davis}},\ }%
  \bibfield{journal}{%
  \bibinfo {journal} {Science}\ }%
  \textbf{\bibinfo {volume} {315}},\ \bibinfo {pages} {1699} (\bibinfo {year}
  {2007})%
  \bibAnnoteFile{NoStop}{smolyaninov2007magnifying}%
\bibitem{zhang2008superlenses}%
  \BibitemOpen
  \bibfield{author}{%
  \bibinfo {author} {\bibfnamefont{X.}~\bibnamefont{Zhang}}\ and\ \bibinfo
  {author} {\bibfnamefont{Z.}~\bibnamefont{Liu}},\ }%
  \bibfield{journal}{%
  \bibinfo {journal} {Nature materials}\ }%
  \textbf{\bibinfo {volume} {7}},\ \bibinfo {pages} {435} (\bibinfo {year}
  {2008})%
  \bibAnnoteFile{NoStop}{zhang2008superlenses}%
\bibitem{mattiucci09}%
  \BibitemOpen
  \bibfield{author}{%
  \bibinfo {author} {\bibfnamefont{N.}~\bibnamefont{Mattiucci}}, \bibinfo
  {author} {\bibfnamefont{G.}~\bibnamefont{D'Aguanno}}, \bibinfo {author}
  {\bibfnamefont{M.}~\bibnamefont{Scalora}}, \bibinfo {author}
  {\bibfnamefont{M.}~\bibnamefont{Bloemer}},\ and\ \bibinfo {author}
  {\bibfnamefont{C.}~\bibnamefont{Sibilia}},\ }%
  \bibfield{journal}{%
  \bibinfo {journal} {Optics express}\ }%
  \textbf{\bibinfo {volume} {17}},\ \bibinfo {pages} {17517} (\bibinfo {year}
  {2009})%
  \bibAnnoteFile{NoStop}{mattiucci09}%
\bibitem{benedicto12}%
  \BibitemOpen
  \bibfield{author}{%
  \bibinfo {author} {\bibfnamefont{J.}~\bibnamefont{B{\'e}n{\'e}dicto}},
  \bibinfo {author} {\bibfnamefont{E.}~\bibnamefont{Centeno}},\ and\ \bibinfo
  {author} {\bibfnamefont{A.}~\bibnamefont{Moreau}},\ }%
  \bibfield{journal}{%
  \bibinfo {journal} {Optics letters}\ }%
  \textbf{\bibinfo {volume} {37}},\ \bibinfo {pages} {4786} (\bibinfo {year}
  {2012})%
  \bibAnnoteFile{NoStop}{benedicto12}%
\bibitem{ciraci12}%
  \BibitemOpen
  \bibfield{author}{%
  \bibinfo {author} {\bibfnamefont{C.}~\bibnamefont{Cirac{\`\i}}}, \bibinfo
  {author} {\bibfnamefont{R.}~\bibnamefont{Hill}}, \bibinfo {author}
  {\bibfnamefont{J.}~\bibnamefont{Mock}}, \bibinfo {author}
  {\bibfnamefont{Y.}~\bibnamefont{Urzhumov}}, \bibinfo {author}
  {\bibfnamefont{A.}~\bibnamefont{Fern{\'a}ndez-Dom{\'\i}nguez}}, \bibinfo
  {author} {\bibfnamefont{S.}~\bibnamefont{Maier}}, \bibinfo {author}
  {\bibfnamefont{J.}~\bibnamefont{Pendry}}, \bibinfo {author}
  {\bibfnamefont{A.}~\bibnamefont{Chilkoti}},\ and\ \bibinfo {author}
  {\bibfnamefont{D.}~\bibnamefont{Smith}},\ }%
  \bibfield{journal}{%
  \bibinfo {journal} {Science}\ }%
  \textbf{\bibinfo {volume} {337}},\ \bibinfo {pages} {1072} (\bibinfo {year}
  {2012})%
  \bibAnnoteFile{NoStop}{ciraci12}%
\bibitem{ruppin05}%
  \BibitemOpen
  \bibfield{author}{%
  \bibinfo {author} {\bibfnamefont{R.}~\bibnamefont{Ruppin}},\ }%
  \bibfield{journal}{%
  \bibinfo {journal} {Journal of Physics: Condensed Matter}\ }%
  \textbf{\bibinfo {volume} {17}},\ \bibinfo {pages} {1803} (\bibinfo {year}
  {2005})%
  \bibAnnoteFile{NoStop}{ruppin05}%
\bibitem{ruppin05b}%
  \BibitemOpen
  \bibfield{author}{%
  \bibinfo {author} {\bibfnamefont{R.}~\bibnamefont{Ruppin}}\ and\ \bibinfo
  {author} {\bibfnamefont{K.}~\bibnamefont{Kempa}},\ }%
  \bibfield{journal}{%
  \bibinfo {journal} {Physical Review B}\ }%
  \textbf{\bibinfo {volume} {72}},\ \bibinfo {pages} {153105} (\bibinfo {year}
  {2005})%
  \bibAnnoteFile{NoStop}{ruppin05b}%
\bibitem{yan13}%
  \BibitemOpen
  \bibfield{author}{%
  \bibinfo {author} {\bibfnamefont{W.}~\bibnamefont{Yan}}, \bibinfo {author}
  {\bibfnamefont{N.}~\bibnamefont{Asger~Mortensen}},\ and\ \bibinfo {author}
  {\bibfnamefont{M.}~\bibnamefont{Wubs}},\ }%
  \bibfield{journal}{%
  \bibinfo {journal} {Optics Express}\ }%
  \textbf{\bibinfo {volume} {21}},\ \bibinfo {pages} {15026} (\bibinfo {year}
  {2013})%
  \bibAnnoteFile{NoStop}{yan13}%
\bibitem{boardman82}%
  \BibitemOpen
  \bibfield{author}{%
  \bibinfo {author} {\bibfnamefont{A.~D.}\ \bibnamefont{Boardman}},\ }%
  \emph{\bibinfo {title} {Electromagnetic surface modes}}\ (\bibinfo
  {publisher} {Wiley},\ \bibinfo {year} {1982})%
  \bibAnnoteFile{NoStop}{boardman82}%
\bibitem{forstmann86}%
  \BibitemOpen
  \bibfield{author}{%
  \bibinfo {author} {\bibfnamefont{F.}~\bibnamefont{Frostmann}}\ and\ \bibinfo
  {author} {\bibfnamefont{R.~R.}\ \bibnamefont{Gerhardts}},\ }%
  \emph{\bibinfo {title} {Metal optics near the plasma frequency}},\ Vol.\
  \bibinfo {volume} {109}\ (\bibinfo {publisher} {Springer-Verlag},\ \bibinfo
  {year} {1986})%
  \bibAnnoteFile{NoStop}{forstmann86}%
\bibitem{moreau13}%
  \BibitemOpen
  \bibfield{author}{%
  \bibinfo {author} {\bibfnamefont{A.}~\bibnamefont{Moreau}}, \bibinfo {author}
  {\bibfnamefont{C.}~\bibnamefont{Cirac{\`\i}}},\ and\ \bibinfo {author}
  {\bibfnamefont{D.~R.}\ \bibnamefont{Smith}},\ }%
  \bibfield{journal}{%
  \bibinfo {journal} {Physical Review B}\ }%
  \textbf{\bibinfo {volume} {87}},\ \bibinfo {pages} {045401} (\bibinfo {year}
  {2013})%
  \bibAnnoteFile{NoStop}{moreau13}%
\bibitem{mochan87}%
  \BibitemOpen
  \bibfield{author}{%
  \bibinfo {author} {\bibfnamefont{W.~L.}\ \bibnamefont{Moch{\'a}n}}, \bibinfo
  {author} {\bibfnamefont{M.}~\bibnamefont{del CastilloMussot}}, \bibinfo
  {author} {\bibfnamefont{R.~G.}\ \bibnamefont{Barrera}}, \bibinfo {author}
  {\bibfnamefont{D.}~\bibnamefont{Federal}}, \emph{et~al.},\ }%
  \bibfield{journal}{%
  \bibinfo {journal} {Physical Review B}\ }%
  \textbf{\bibinfo {volume} {35}},\ \bibinfo {pages} {1088} (\bibinfo {year}
  {1987})%
  \bibAnnoteFile{NoStop}{mochan87}%
\bibitem{yan12}%
  \BibitemOpen
  \bibfield{author}{%
  \bibinfo {author} {\bibfnamefont{W.}~\bibnamefont{Yan}}, \bibinfo {author}
  {\bibfnamefont{M.}~\bibnamefont{Wubs}},\ and\ \bibinfo {author}
  {\bibfnamefont{N.~A.}\ \bibnamefont{Mortensen}},\ }%
  \bibfield{journal}{%
  \bibinfo {journal} {Physical Review B}\ }%
  \textbf{\bibinfo {volume} {86}},\ \bibinfo {pages} {205429} (\bibinfo {year}
  {2012})%
  \bibAnnoteFile{NoStop}{yan12}%
\bibitem{scalora10}%
  \BibitemOpen
  \bibfield{author}{%
  \bibinfo {author} {\bibfnamefont{M.}~\bibnamefont{Scalora}}, \bibinfo
  {author} {\bibfnamefont{M.~A.}\ \bibnamefont{Vincenti}}, \bibinfo {author}
  {\bibfnamefont{D.}~\bibnamefont{de~Ceglia}}, \bibinfo {author}
  {\bibfnamefont{V.}~\bibnamefont{Roppo}}, \bibinfo {author}
  {\bibfnamefont{M.}~\bibnamefont{Centini}}, \bibinfo {author}
  {\bibfnamefont{N.}~\bibnamefont{Akozbek}},\ and\ \bibinfo {author}
  {\bibfnamefont{M.~J.}\ \bibnamefont{Bloemer}},\ }%
  \bibfield{journal}{%
  \bibinfo {journal} {Phys. Rev. A}\ }%
  \textbf{\bibinfo {volume} {82}},\ \bibinfo {pages} {043828} (\bibinfo {year}
  {2010})%
  \bibAnnoteFile{NoStop}{scalora10}%
\bibitem{crouseilles08}%
  \BibitemOpen
  \bibfield{author}{%
  \bibinfo {author} {\bibfnamefont{N.}~\bibnamefont{Crouseilles}}, \bibinfo
  {author} {\bibfnamefont{P.~A.}\ \bibnamefont{Hervieux}},\ and\ \bibinfo
  {author} {\bibfnamefont{G.}~\bibnamefont{Manfredi}},\ }%
  \bibfield{journal}{%
  \bibinfo {journal} {Phys. Rev. B}\ }%
  \textbf{\bibinfo {volume} {78}},\ \bibinfo {pages} {155412} (\bibinfo {year}
  {2008})%
  \bibAnnoteFile{NoStop}{crouseilles08}%
\bibitem{kliewer68}%
  \BibitemOpen
  \bibfield{author}{%
  \bibinfo {author} {\bibfnamefont{K.}~\bibnamefont{Kliewer}}\ and\ \bibinfo
  {author} {\bibfnamefont{R.}~\bibnamefont{Fuchs}},\ }%
  \bibfield{journal}{%
  \bibinfo {journal} {Physical Review}\ }%
  \textbf{\bibinfo {volume} {172}},\ \bibinfo {pages} {607} (\bibinfo {year}
  {1968})%
  \bibAnnoteFile{NoStop}{kliewer68}%
\bibitem{melnyk70}%
  \BibitemOpen
  \bibfield{author}{%
  \bibinfo {author} {\bibfnamefont{A.~R.}\ \bibnamefont{Melnyk}}\ and\ \bibinfo
  {author} {\bibfnamefont{M.~J.}\ \bibnamefont{Harrison}},\ }%
  \bibfield{journal}{%
  \bibinfo {journal} {Phys. Rev. B}\ }%
  \textbf{\bibinfo {volume} {2}},\ \bibinfo {pages} {835} (\bibinfo {year}
  {1970})%
  \bibAnnoteFile{NoStop}{melnyk70}%
\bibitem{fuchs81}%
  \BibitemOpen
  \bibfield{author}{%
  \bibinfo {author} {\bibfnamefont{R.}~\bibnamefont{Fuchs}}\ and\ \bibinfo
  {author} {\bibfnamefont{R.~G.}\ \bibnamefont{Barrera}},\ }%
  \bibfield{journal}{%
  \bibinfo {journal} {Physical Review B}\ }%
  \textbf{\bibinfo {volume} {24}},\ \bibinfo {pages} {2940} (\bibinfo {year}
  {1981})%
  \bibAnnoteFile{NoStop}{fuchs81}%
\bibitem{gerhardts84}%
  \BibitemOpen
  \bibfield{author}{%
  \bibinfo {author} {\bibfnamefont{R.~R.}\ \bibnamefont{Gerhardts}}\ and\
  \bibinfo {author} {\bibfnamefont{K.}~\bibnamefont{Kempa}},\ }%
  \bibfield{journal}{%
  \bibinfo {journal} {Physical Review B}\ }%
  \textbf{\bibinfo {volume} {30}},\ \bibinfo {pages} {5704} (\bibinfo {year}
  {1984})%
  \bibAnnoteFile{NoStop}{gerhardts84}%
\bibitem{agarwal83}%
  \BibitemOpen
  \bibfield{author}{%
  \bibinfo {author} {\bibfnamefont{G.~S.}\ \bibnamefont{Agarwal}}\ and\
  \bibinfo {author} {\bibfnamefont{S.~V.}\ \bibnamefont{ONeil}},\ }%
  \bibfield{journal}{%
  \bibinfo {journal} {Physical Review B}\ }%
  \textbf{\bibinfo {volume} {28}},\ \bibinfo {pages} {487} (\bibinfo {year}
  {1983})%
  \bibAnnoteFile{NoStop}{agarwal83}%
\bibitem{fuchs87}%
  \BibitemOpen
  \bibfield{author}{%
  \bibinfo {author} {\bibfnamefont{R.}~\bibnamefont{Fuchs}}\ and\ \bibinfo
  {author} {\bibfnamefont{F.}~\bibnamefont{Claro}},\ }%
  \bibfield{journal}{%
  \bibinfo {journal} {Physical Review B}\ }%
  \textbf{\bibinfo {volume} {35}},\ \bibinfo {pages} {3722} (\bibinfo {year}
  {1987})%
  \bibAnnoteFile{NoStop}{fuchs87}%
\bibitem{halevi95}%
  \BibitemOpen
  \bibfield{author}{%
  \bibinfo {author} {\bibfnamefont{P.}~\bibnamefont{Halevi}},\ }%
  \bibfield{journal}{%
  \bibinfo {journal} {Physical Review B}\ }%
  \textbf{\bibinfo {volume} {51}},\ \bibinfo {pages} {7497} (\bibinfo {year}
  {1995})%
  \bibAnnoteFile{NoStop}{halevi95}%
\bibitem{ford84}%
  \BibitemOpen
  \bibfield{author}{%
  \bibinfo {author} {\bibfnamefont{G.~W.}\ \bibnamefont{Ford}}\ and\ \bibinfo
  {author} {\bibfnamefont{W.}~\bibnamefont{Weber}},\ }%
  \bibfield{journal}{%
  \bibinfo {journal} {Physics Reports}\ }%
  \textbf{\bibinfo {volume} {113}},\ \bibinfo {pages} {195} (\bibinfo {year}
  {1984})%
  \bibAnnoteFile{NoStop}{ford84}%
\bibitem{chapuis08}%
  \BibitemOpen
  \bibfield{author}{%
  \bibinfo {author} {\bibfnamefont{P.-O.}\ \bibnamefont{Chapuis}}, \bibinfo
  {author} {\bibfnamefont{S.}~\bibnamefont{Volz}}, \bibinfo {author}
  {\bibfnamefont{C.}~\bibnamefont{Henkel}}, \bibinfo {author}
  {\bibfnamefont{K.}~\bibnamefont{Joulain}},\ and\ \bibinfo {author}
  {\bibfnamefont{J.-J.}\ \bibnamefont{Greffet}},\ }%
  \bibfield{journal}{%
  \bibinfo {journal} {Physical Review B}\ }%
  \textbf{\bibinfo {volume} {77}},\ \bibinfo {pages} {035431} (\bibinfo {year}
  {2008})%
  \bibAnnoteFile{NoStop}{chapuis08}%
\bibitem{feibelman82}%
  \BibitemOpen
  \bibfield{author}{%
  \bibinfo {author} {\bibfnamefont{P.}~\bibnamefont{Feibelman}},\ }%
  \bibfield{journal}{%
  \bibinfo {journal} {Progress in Surface Science}\ }%
  \textbf{\bibinfo {volume} {12}},\ \bibinfo {pages} {287} (\bibinfo {year}
  {1982})%
  \bibAnnoteFile{NoStop}{feibelman82}%
\bibitem{liebsch1987dynamical}%
  \BibitemOpen
  \bibfield{author}{%
  \bibinfo {author} {\bibfnamefont{A.}~\bibnamefont{Liebsch}},\ }%
  \bibfield{journal}{%
  \bibinfo {journal} {Physical Review B}\ }%
  \textbf{\bibinfo {volume} {36}},\ \bibinfo {pages} {7378} (\bibinfo {year}
  {1987})%
  \bibAnnoteFile{NoStop}{liebsch1987dynamical}%
\bibitem{liebsch1995influence}%
  \BibitemOpen
  \bibfield{author}{%
  \bibinfo {author} {\bibfnamefont{A.}~\bibnamefont{Liebsch}}\ and\ \bibinfo
  {author} {\bibfnamefont{W.~L.}\ \bibnamefont{Schaich}},\ }%
  \bibfield{journal}{%
  \bibinfo {journal} {Physical Review B}\ }%
  \textbf{\bibinfo {volume} {52}},\ \bibinfo {pages} {14219} (\bibinfo {year}
  {1995})%
  \bibAnnoteFile{NoStop}{liebsch1995influence}%
\bibitem{savage12}%
  \BibitemOpen
  \bibfield{author}{%
  \bibinfo {author} {\bibfnamefont{K.~J.}\ \bibnamefont{Savage}}, \bibinfo
  {author} {\bibfnamefont{M.~M.}\ \bibnamefont{Hawkeye}}, \bibinfo {author}
  {\bibfnamefont{R.}~\bibnamefont{Esteban}}, \bibinfo {author}
  {\bibfnamefont{A.~G.}\ \bibnamefont{Borisov}}, \bibinfo {author}
  {\bibfnamefont{J.}~\bibnamefont{Aizpurua}},\ and\ \bibinfo {author}
  {\bibfnamefont{J.~J.}\ \bibnamefont{Baumberg}},\ }%
  \bibfield{journal}{%
  \bibinfo {journal} {Nature}\ }%
  \textbf{\bibinfo {volume} {491}},\ \bibinfo {pages} {574} (\bibinfo {year}
  {2012})%
  \bibAnnoteFile{NoStop}{savage12}%
\bibitem{teperik13}%
  \BibitemOpen
  \bibfield{author}{%
  \bibinfo {author} {\bibfnamefont{T.}~\bibnamefont{Teperik}}, \bibinfo
  {author} {\bibfnamefont{P.}~\bibnamefont{Nordlander}}, \bibinfo {author}
  {\bibfnamefont{J.}~\bibnamefont{Aizpurua}},\ and\ \bibinfo {author}
  {\bibfnamefont{A.}~\bibnamefont{Borisov}},\ }%
  \bibfield{journal}{%
  \bibinfo {journal} {Physical review letters}\ }%
  \textbf{\bibinfo {volume} {110}},\ \bibinfo {pages} {263901} (\bibinfo {year}
  {2013})%
  \bibAnnoteFile{NoStop}{teperik13}%
\bibitem{esteban12}%
  \BibitemOpen
  \bibfield{author}{%
  \bibinfo {author} {\bibfnamefont{R.}~\bibnamefont{Esteban}}, \bibinfo
  {author} {\bibfnamefont{A.~G.}\ \bibnamefont{Borisov}}, \bibinfo {author}
  {\bibfnamefont{P.}~\bibnamefont{Nordlander}},\ and\ \bibinfo {author}
  {\bibfnamefont{J.}~\bibnamefont{Aizpurua}},\ }%
  \bibfield{journal}{%
  \bibinfo {journal} {Nature Communications}\ }%
  \textbf{\bibinfo {volume} {3}},\ \bibinfo {pages} {825} (\bibinfo {year}
  {2012})%
  \bibAnnoteFile{NoStop}{esteban12}%
\bibitem{ciraci2013hydrodynamic}%
  \BibitemOpen
  \bibfield{author}{%
  \bibinfo {author} {\bibfnamefont{C.}~\bibnamefont{Cirac{\`\i}}}, \bibinfo
  {author} {\bibfnamefont{J.~B.}\ \bibnamefont{Pendry}},\ and\ \bibinfo
  {author} {\bibfnamefont{D.~R.}\ \bibnamefont{Smith}},\ }%
  \bibfield{journal}{%
  \bibinfo {journal} {ChemPhysChem}\ }%
  \textbf{\bibinfo {volume} {14}},\ \bibinfo {pages} {1109} (\bibinfo {year}
  {2013})%
  \bibAnnoteFile{NoStop}{ciraci2013hydrodynamic}%
\bibitem{luo13}%
  \BibitemOpen
  \bibfield{author}{%
  \bibinfo {author} {\bibfnamefont{Y.}~\bibnamefont{Luo}}, \bibinfo {author}
  {\bibfnamefont{A.}~\bibnamefont{Fernandez-Dominguez}}, \bibinfo {author}
  {\bibfnamefont{A.}~\bibnamefont{Wiener}}, \bibinfo {author}
  {\bibfnamefont{S.~A.}\ \bibnamefont{Maier}},\ and\ \bibinfo {author}
  {\bibfnamefont{J.}~\bibnamefont{Pendry}},\ }%
  \bibfield{journal}{%
  \bibinfo {journal} {Physical review letters}\ }%
  \textbf{\bibinfo {volume} {111}},\ \bibinfo {pages} {093901} (\bibinfo {year}
  {2013})%
  \bibAnnoteFile{NoStop}{luo13}%
\bibitem{david13}%
  \BibitemOpen
  \bibfield{author}{%
  \bibinfo {author} {\bibfnamefont{C.}~\bibnamefont{David}}, \bibinfo {author}
  {\bibfnamefont{N.~A.}\ \bibnamefont{Mortensen}},\ and\ \bibinfo {author}
  {\bibfnamefont{J.}~\bibnamefont{Christensen}},\ }%
  \bibfield{journal}{%
  \bibinfo {journal} {Scientific reports}\ }%
  \textbf{\bibinfo {volume} {3}} (\bibinfo {year} {2013})%
  \bibAnnoteFile{NoStop}{david13}%
\bibitem{scholl12}%
  \BibitemOpen
  \bibfield{author}{%
  \bibinfo {author} {\bibfnamefont{J.~A.}\ \bibnamefont{Scholl}}, \bibinfo
  {author} {\bibfnamefont{A.~L.}\ \bibnamefont{Koh}},\ and\ \bibinfo {author}
  {\bibfnamefont{J.~A.}\ \bibnamefont{Dionne}},\ }%
  \bibfield{journal}{%
  \bibinfo {journal} {Nature}\ }%
  \textbf{\bibinfo {volume} {483}},\ \bibinfo {pages} {421} (\bibinfo {year}
  {2012})%
  \bibAnnoteFile{NoStop}{scholl12}%
\bibitem{scholl13}%
  \BibitemOpen
  \bibfield{author}{%
  \bibinfo {author} {\bibfnamefont{J.~A.}\ \bibnamefont{Scholl}}, \bibinfo
  {author} {\bibfnamefont{A.}~\bibnamefont{Garc{\'\i}a-Etxarri}}, \bibinfo
  {author} {\bibfnamefont{A.~L.}\ \bibnamefont{Koh}},\ and\ \bibinfo {author}
  {\bibfnamefont{J.~A.}\ \bibnamefont{Dionne}},\ }%
  \bibfield{journal}{%
  \bibinfo {journal} {Nano letters}\ }%
  \textbf{\bibinfo {volume} {13}},\ \bibinfo {pages} {564} (\bibinfo {year}
  {2013})%
  \bibAnnoteFile{NoStop}{scholl13}%
\bibitem{fernandez12}%
  \BibitemOpen
  \bibfield{author}{%
  \bibinfo {author} {\bibfnamefont{A.~I.}\ \bibnamefont{Fernandez-Dominguez}},
  \bibinfo {author} {\bibfnamefont{A.}~\bibnamefont{Wiener}}, \bibinfo {author}
  {\bibfnamefont{F.~J.}\ \bibnamefont{Garcia-Vidal}}, \bibinfo {author}
  {\bibfnamefont{S.~A.}\ \bibnamefont{Maier}},\ and\ \bibinfo {author}
  {\bibfnamefont{J.~B.}\ \bibnamefont{Pendry}},\ }%
  \bibfield{journal}{%
  \bibinfo {journal} {Physical Review Letters}\ }%
  \textbf{\bibinfo {volume} {108}},\ \bibinfo {pages} {106802} (\bibinfo {year}
  {2012})%
  \bibAnnoteFile{NoStop}{fernandez12}%
\bibitem{ciraci2013effects}%
  \BibitemOpen
  \bibfield{author}{%
  \bibinfo {author} {\bibfnamefont{C.}~\bibnamefont{Cirac{\`\i}}}, \bibinfo
  {author} {\bibfnamefont{Y.}~\bibnamefont{Urzhumov}},\ and\ \bibinfo {author}
  {\bibfnamefont{D.~R.}\ \bibnamefont{Smith}},\ }%
  \bibfield{journal}{%
  \bibinfo {journal} {JOSA B}\ }%
  \textbf{\bibinfo {volume} {30}},\ \bibinfo {pages} {2731} (\bibinfo {year}
  {2013})%
  \bibAnnoteFile{NoStop}{ciraci2013effects}%
\bibitem{wiener12}%
  \BibitemOpen
  \bibfield{author}{%
  \bibinfo {author} {\bibfnamefont{A.}~\bibnamefont{Wiener}}, \bibinfo {author}
  {\bibfnamefont{A.~I.}\ \bibnamefont{Fernández-Domínguez}}, \bibinfo
  {author} {\bibfnamefont{A.~P.}\ \bibnamefont{Horsfield}}, \bibinfo {author}
  {\bibfnamefont{J.~B.}\ \bibnamefont{Pendry}},\ and\ \bibinfo {author}
  {\bibfnamefont{S.~A.}\ \bibnamefont{Maier}},\ }%
  \bibfield{journal}{%
  \Doi{10.1021/nl301478n}{\bibinfo {journal} {Nano Letters}}\ }%
  \textbf{\bibinfo {volume} {12}},\ \bibinfo {pages} {3308} (\bibinfo {year}
  {2012}),\
  \Eprint{http://arxiv.org/abs/http://pubs.acs.org/doi/pdf/10.1021/nl301478n}{%
http://pubs.acs.org/doi/pdf/10.1021/nl301478n},\
  \url{http://pubs.acs.org/doi/abs/10.1021/nl301478n}%
  \bibAnnoteFile{NoStop}{wiener12}%
\bibitem{fernandez12a}%
  \BibitemOpen
  \bibfield{author}{%
  \bibinfo {author} {\bibfnamefont{A.}~\bibnamefont{Fernandez-Dominguez}},
  \bibinfo {author} {\bibfnamefont{P.}~\bibnamefont{Zhang}}, \bibinfo {author}
  {\bibfnamefont{Y.}~\bibnamefont{Luo}}, \bibinfo {author}
  {\bibfnamefont{S.}~\bibnamefont{Maier}}, \bibinfo {author}
  {\bibfnamefont{F.}~\bibnamefont{Garcia-Vidal}},\ and\ \bibinfo {author}
  {\bibfnamefont{J.}~\bibnamefont{Pendry}},\ }%
  \bibfield{journal}{%
  \bibinfo {journal} {Physical Review B}\ }%
  \textbf{\bibinfo {volume} {86}},\ \bibinfo {pages} {241110} (\bibinfo {year}
  {2012})%
  \bibAnnoteFile{NoStop}{fernandez12a}%
\bibitem{raza13}%
  \BibitemOpen
  \bibfield{author}{%
  \bibinfo {author} {\bibfnamefont{S.}~\bibnamefont{Raza}}, \bibinfo {author}
  {\bibfnamefont{T.}~\bibnamefont{Christensen}}, \bibinfo {author}
  {\bibfnamefont{M.}~\bibnamefont{Wubs}}, \bibinfo {author}
  {\bibfnamefont{S.}~\bibnamefont{Bozhevolnyi}},\ and\ \bibinfo {author}
  {\bibfnamefont{N.}~\bibnamefont{Mortensen}},\ }%
  \bibfield{journal}{%
  \bibinfo {journal} {Physical Review B (Condensed Matter and Materials
  Physics)}\ }%
  \textbf{\bibinfo {volume} {88}} (\bibinfo {year} {2013}),\ ISSN \bibinfo
  {issn} {1098-0121}%
  \bibAnnoteFile{NoStop}{raza13}%
\bibitem{moreau12b}%
  \BibitemOpen
  \bibfield{author}{%
  \bibinfo {author} {\bibfnamefont{A.}~\bibnamefont{Moreau}}, \bibinfo {author}
  {\bibfnamefont{C.}~\bibnamefont{Cirac{\`\i}}}, \bibinfo {author}
  {\bibfnamefont{J.~J.}\ \bibnamefont{Mock}}, \bibinfo {author}
  {\bibfnamefont{R.~T.}\ \bibnamefont{Hill}}, \bibinfo {author}
  {\bibfnamefont{Q.}~\bibnamefont{Wang}}, \bibinfo {author}
  {\bibfnamefont{B.~J.}\ \bibnamefont{Wiley}}, \bibinfo {author}
  {\bibfnamefont{A.}~\bibnamefont{Chilkoti}},\ and\ \bibinfo {author}
  {\bibfnamefont{D.~R.}\ \bibnamefont{Smith}},\ }%
  \bibfield{journal}{%
  \bibinfo {journal} {Nature}\ }%
  \textbf{\bibinfo {volume} {492}},\ \bibinfo {pages} {86} (\bibinfo {year}
  {2012})%
  \bibAnnoteFile{NoStop}{moreau12b}%
\bibitem{raza11}%
  \BibitemOpen
  \bibfield{author}{%
  \bibinfo {author} {\bibfnamefont{S.}~\bibnamefont{Raza}}, \bibinfo {author}
  {\bibfnamefont{G.}~\bibnamefont{Toscano}}, \bibinfo {author}
  {\bibfnamefont{A.~P.}\ \bibnamefont{Jauho}}, \bibinfo {author}
  {\bibfnamefont{M.}~\bibnamefont{Wubs}},\ and\ \bibinfo {author}
  {\bibfnamefont{N.~A.}\ \bibnamefont{Mortensen}},\ }%
  \bibfield{journal}{%
  \bibinfo {journal} {Physical Review B}\ }%
  \textbf{\bibinfo {volume} {84}},\ \bibinfo {pages} {121412} (\bibinfo {year}
  {2011})%
  \bibAnnoteFile{NoStop}{raza11}%
\bibitem{ciraci13}%
  \BibitemOpen
  \bibfield{author}{%
  \bibinfo {author} {\bibfnamefont{C.}~\bibnamefont{Cirac{\`\i}}}, \bibinfo
  {author} {\bibfnamefont{J.~B.}\ \bibnamefont{Pendry}},\ and\ \bibinfo
  {author} {\bibfnamefont{D.~R.}\ \bibnamefont{Smith}},\ }%
  \bibfield{journal}{%
  \bibinfo {journal} {ChemPhysChem}\ }%
  \textbf{\bibinfo {volume} {14}},\ \bibinfo {pages} {1109} (\bibinfo {year}
  {2013})%
  \bibAnnoteFile{NoStop}{ciraci13}%
\bibitem{rakic98}%
  \BibitemOpen
  \bibfield{author}{%
  \bibinfo {author} {\bibfnamefont{A.~D.}\ \bibnamefont{Rakic}}, \bibinfo
  {author} {\bibfnamefont{A.~B.}\ \bibnamefont{Djuri{\v{s}}ic}}, \bibinfo
  {author} {\bibfnamefont{J.~M.}\ \bibnamefont{Elazar}},\ and\ \bibinfo
  {author} {\bibfnamefont{M.~L.}\ \bibnamefont{Majewski}},\ }%
  \bibfield{journal}{%
  \bibinfo {journal} {Applied Optics}\ }%
  \textbf{\bibinfo {volume} {37}},\ \bibinfo {pages} {5271} (\bibinfo {year}
  {1998})%
  \bibAnnoteFile{NoStop}{rakic98}%
\bibitem{mochan88}%
  \BibitemOpen
  \bibfield{author}{%
  \bibinfo {author} {\bibfnamefont{W.~L.}\ \bibnamefont{Moch{\'a}n}}\ and\
  \bibinfo {author} {\bibfnamefont{M.}~\bibnamefont{del CastilloMussot}},\ }%
  \bibfield{journal}{%
  \bibinfo {journal} {Physical Review B}\ }%
  \textbf{\bibinfo {volume} {37}},\ \bibinfo {pages} {6763} (\bibinfo {year}
  {1988})%
  \bibAnnoteFile{NoStop}{mochan88}%
\bibitem{smith03}%
  \BibitemOpen
  \bibfield{author}{%
  \bibinfo {author} {\bibfnamefont{D.}~\bibnamefont{Smith}}\ and\ \bibinfo
  {author} {\bibfnamefont{D.}~\bibnamefont{Schurig}},\ }%
  \bibfield{journal}{%
  \bibinfo {journal} {Physical Review Letters}\ }%
  \textbf{\bibinfo {volume} {90}},\ \bibinfo {pages} {077405} (\bibinfo {year}
  {2003})%
  \bibAnnoteFile{NoStop}{smith03}%
\bibitem{smith04}%
  \BibitemOpen
  \bibfield{author}{%
  \bibinfo {author} {\bibfnamefont{D.~R.}\ \bibnamefont{Smith}}, \bibinfo
  {author} {\bibfnamefont{P.}~\bibnamefont{Kolinko}},\ and\ \bibinfo {author}
  {\bibfnamefont{D.}~\bibnamefont{Schurig}},\ }%
  \bibfield{journal}{%
  \bibinfo {journal} {JOSA B}\ }%
  \textbf{\bibinfo {volume} {21}},\ \bibinfo {pages} {1032} (\bibinfo {year}
  {2004})%
  \bibAnnoteFile{NoStop}{smith04}%
\bibitem{bloemer07}%
  \BibitemOpen
  \bibfield{author}{%
  \bibinfo {author} {\bibfnamefont{M.}~\bibnamefont{Bloemer}}, \bibinfo
  {author} {\bibfnamefont{G.}~\bibnamefont{DAguanno}}, \bibinfo {author}
  {\bibfnamefont{N.}~\bibnamefont{Mattiucci}}, \bibinfo {author}
  {\bibfnamefont{M.}~\bibnamefont{Scalora}},\ and\ \bibinfo {author}
  {\bibfnamefont{N.}~\bibnamefont{Akozbek}},\ }%
  \bibfield{journal}{%
  \bibinfo {journal} {Applied physics letters}\ }%
  \textbf{\bibinfo {volume} {90}},\ \bibinfo {pages} {174113} (\bibinfo {year}
  {2007})%
  \bibAnnoteFile{NoStop}{bloemer07}%
\bibitem{benedicto13}%
  \BibitemOpen
  \bibfield{author}{%
  \bibinfo {author} {\bibfnamefont{J.}~\bibnamefont{B{\'e}n{\'e}dicto}},
  \bibinfo {author} {\bibfnamefont{E.}~\bibnamefont{Centeno}}, \bibinfo
  {author} {\bibfnamefont{R.}~\bibnamefont{Poll{\`e}s}},\ and\ \bibinfo
  {author} {\bibfnamefont{A.}~\bibnamefont{Moreau}},\ }%
  \bibfield{journal}{%
  \bibinfo {journal} {Phys. Rev. B}\ }%
  \textbf{\bibinfo {volume} {88}},\ \bibinfo {pages} {245138} (\bibinfo {year}
  {2013})%
  \bibAnnoteFile{NoStop}{benedicto13}%
\bibitem{krayzel10}%
  \BibitemOpen
  \bibfield{author}{%
  \bibinfo {author} {\bibfnamefont{F.}~\bibnamefont{Krayzel}}, \bibinfo
  {author} {\bibfnamefont{R.}~\bibnamefont{Polles}}, \bibinfo {author}
  {\bibfnamefont{A.}~\bibnamefont{Moreau}}, \bibinfo {author}
  {\bibfnamefont{M.}~\bibnamefont{Mihailovic}},\ and\ \bibinfo {author}
  {\bibfnamefont{G.}~\bibnamefont{Granet}},\ }%
  \bibfield{journal}{%
  \bibinfo {journal} {Journal of the European Optical Society-Rapid
  Publications}\ }%
  \textbf{\bibinfo {volume} {5}},\ \bibinfo {pages} {10025} (\bibinfo {year}
  {2010}),\ \url{http://elma.univ-bpclermont.fr/moreau/moosh/index_en.html}%
  \bibAnnoteFile{NoStop}{krayzel10}%
\bibitem{abeles50}%
  \BibitemOpen
  \bibfield{author}{%
  \bibinfo {author} {\bibfnamefont{F.}~\bibnamefont{Abel{\`e}s}},\ }%
  \bibfield{journal}{%
  \bibinfo {journal} {J. Phys. Radium}\ }%
  \textbf{\bibinfo {volume} {11}},\ \bibinfo {pages} {307} (\bibinfo {year}
  {1950})%
  \bibAnnoteFile{NoStop}{abeles50}%
\bibitem{markos08}%
  \BibitemOpen
  \bibfield{author}{%
  \bibinfo {author} {\bibfnamefont{P.}~\bibnamefont{Markos}}\ and\ \bibinfo
  {author} {\bibfnamefont{C.~M.}\ \bibnamefont{Soukoulis}},\ }%
  \emph{\bibinfo {title} {Wave propagation: from electrons to photonic crystals
  and left-handed materials}}\ (\bibinfo {publisher} {Princeton University
  Press},\ \bibinfo {year} {2008})%
  \bibAnnoteFile{NoStop}{markos08}%
\bibitem{verhagen10}%
  \BibitemOpen
  \bibfield{author}{%
  \bibinfo {author} {\bibfnamefont{E.}~\bibnamefont{Verhagen}}, \bibinfo
  {author} {\bibfnamefont{R.}~\bibnamefont{de~Waele}}, \bibinfo {author}
  {\bibfnamefont{L.}~\bibnamefont{Kuipers}},\ and\ \bibinfo {author}
  {\bibfnamefont{A.}~\bibnamefont{Polman}},\ }%
  \bibfield{journal}{%
  \bibinfo {journal} {Physical review letters}\ }%
  \textbf{\bibinfo {volume} {105}},\ \bibinfo {pages} {223901} (\bibinfo {year}
  {2010})%
  \bibAnnoteFile{NoStop}{verhagen10}%
\bibitem{benedicto11b}%
  \BibitemOpen
  \bibfield{author}{%
  \bibinfo {author} {\bibfnamefont{J.}~\bibnamefont{Benedicto}}, \bibinfo
  {author} {\bibfnamefont{R.}~\bibnamefont{Poll{\`e}s}}, \bibinfo {author}
  {\bibfnamefont{A.}~\bibnamefont{Moreau}},\ and\ \bibinfo {author}
  {\bibfnamefont{E.}~\bibnamefont{Centeno}},\ }%
  \bibfield{journal}{%
  \bibinfo {journal} {Optics letters}\ }%
  \textbf{\bibinfo {volume} {36}},\ \bibinfo {pages} {2539} (\bibinfo {year}
  {2011})%
  \bibAnnoteFile{NoStop}{benedicto11b}%
\bibitem{lalanne96}%
  \BibitemOpen
  \bibfield{author}{%
  \bibinfo {author} {\bibfnamefont{P.}~\bibnamefont{Lalanne}}\ and\ \bibinfo
  {author} {\bibfnamefont{G.~M.}\ \bibnamefont{Morris}},\ }%
  \bibfield{journal}{%
  \bibinfo {journal} {J. Opt. Soc. Am. A}\ }%
  \textbf{\bibinfo {volume} {13}},\ \bibinfo {pages} {779} (\bibinfo {year}
  {1996})%
  \bibAnnoteFile{NoStop}{lalanne96}%
\bibitem{granet96}%
  \BibitemOpen
  \bibfield{author}{%
  \bibinfo {author} {\bibfnamefont{G.}~\bibnamefont{Granet}}\ and\ \bibinfo
  {author} {\bibfnamefont{B.}~\bibnamefont{Guizal}},\ }%
  \bibfield{journal}{%
  \bibinfo {journal} {J. Opt. Soc. Am. A}\ }%
  \textbf{\bibinfo {volume} {13}},\ \bibinfo {pages} {1019} (\bibinfo {year}
  {1996})%
  \bibAnnoteFile{NoStop}{granet96}%
\bibitem{benedicto11a}%
  \BibitemOpen
  \bibfield{author}{%
  \bibinfo {author} {\bibfnamefont{J.}~\bibnamefont{Benedicto}}, \bibinfo
  {author} {\bibfnamefont{A.}~\bibnamefont{Moreau}}, \bibinfo {author}
  {\bibfnamefont{R.}~\bibnamefont{Poll{\`e}s}},\ and\ \bibinfo {author}
  {\bibfnamefont{E.}~\bibnamefont{Centeno}},\ }%
  \bibfield{journal}{%
  \bibinfo {journal} {Solid State Communications}\ }%
  \textbf{\bibinfo {volume} {151}},\ \bibinfo {pages} {354} (\bibinfo {year}
  {2011})%
  \bibAnnoteFile{NoStop}{benedicto11a}%
\bibitem{bozhevolnyi07}%
  \BibitemOpen
  \bibfield{author}{%
  \bibinfo {author} {\bibfnamefont{S.~I.}\ \bibnamefont{Bozhevolnyi}}\ and\
  \bibinfo {author} {\bibfnamefont{T.}~\bibnamefont{S{\o}ndergaard}},\ }%
  \bibfield{journal}{%
  \bibinfo {journal} {Optics Express}\ }%
  \textbf{\bibinfo {volume} {15}},\ \bibinfo {pages} {10869} (\bibinfo {year}
  {2007})%
  \bibAnnoteFile{NoStop}{bozhevolnyi07}%
\bibitem{devore51}%
  \BibitemOpen
  \bibfield{author}{%
  \bibinfo {author} {\bibfnamefont{J.~R.}\ \bibnamefont{DeVore}},\ }%
  \bibfield{journal}{%
  \bibinfo {journal} {J. Opt. Soc. Am.}\ }%
  \textbf{\bibinfo {volume} {41}},\ \bibinfo {pages} {416} (\bibinfo {year}
  {1951})%
  \bibAnnoteFile{NoStop}{devore51}%
\bibitem{pgp}%
  \BibitemOpen
  \url{http://elma.univ-bpclermont.fr/moreau/pgp/index_en.html}%
  \bibAnnoteFile{NoStop}{pgp}%
\bibitem{toscano12}%
  \BibitemOpen
  \bibfield{author}{%
  \bibinfo {author} {\bibfnamefont{G.}~\bibnamefont{Toscano}}, \bibinfo
  {author} {\bibfnamefont{S.}~\bibnamefont{Raza}}, \bibinfo {author}
  {\bibfnamefont{A.-P.}\ \bibnamefont{Jauho}}, \bibinfo {author}
  {\bibfnamefont{N.~A.}\ \bibnamefont{Mortensen}},\ and\ \bibinfo {author}
  {\bibfnamefont{M.}~\bibnamefont{Wubs}},\ }%
  \bibfield{journal}{%
  \bibinfo {journal} {Optics express}\ }%
  \textbf{\bibinfo {volume} {20}},\ \bibinfo {pages} {4176} (\bibinfo {year}
  {2012})%
  \bibAnnoteFile{NoStop}{toscano12}%
\bibitem{toscano14}%
  \BibitemOpen
  \bibfield{author}{%
  \bibinfo {author} {\bibfnamefont{G.}~\bibnamefont{Toscano}}, \bibinfo
  {author} {\bibfnamefont{C.}~\bibnamefont{Rockstuhl}}, \bibinfo {author}
  {\bibfnamefont{F.}~\bibnamefont{Evers}}, \bibinfo {author}
  {\bibfnamefont{H.}~\bibnamefont{Xu}}, \bibinfo {author}
  {\bibfnamefont{N.~A.}\ \bibnamefont{Mortensen}},\ and\ \bibinfo {author}
  {\bibfnamefont{M.}~\bibnamefont{Wubs}},\ }%
  \enquote{\bibinfo {title} {Self-consistent hydrodynamic approach to
  nanoplasmonics: Resonance shifts and spill-out effects},}\  (\bibinfo {year}
  {2014})%
  \bibAnnoteFile{NoStop}{toscano14}%
\end{thebibliography}%

\end{document}